\documentclass[11pt]{article}

 \usepackage{mathrsfs}
 \usepackage[utf8]{inputenc}
 \usepackage{epsfig,graphicx,color}
 \usepackage{latexsym}                 
 \usepackage{amsmath,amsfonts,verbatim,xkeyval,bm, amsthm, upgreek, amscd, wasysym, amssymb}
 \usepackage[english]{babel}
 \usepackage{listings}
 \usepackage{multicol}
 \usepackage[colorlinks]{hyperref}  
 \hypersetup{hidelinks}  
 \usepackage{tikz-cd}
 \usepackage{booktabs}
 \usepackage{cases}
 \usepackage{subfig}
  \usepackage{float}
 \usepackage{makecell}

\newtheorem{Theorem}{Theorem}[section]

\def\avg#1{\left\langle #1 \right\rangle}

\def\vet#1{{\bm #1}}

\def\interi{\mathbb{Z}}

\def\Hscr{\mathcal{H}}

\def\Oscr{\mathcal{O}}

\def\Rscr{\mathcal{R}}

\def\norm#1{||\, #1 \,||}

\def\poisson#1#2{\lbrace #1,#2 \rbrace}
\def\bigpoisson#1#2{\big\{ #1\,,\,#2 \big\}}

\def\q{\tilde{q}}
\def\p{\tilde{p}}

\begin{document}

\pagenumbering{arabic}

\begin{center}
{\Large\bf Kolmogorov algorithm \\for isochronous Hamiltonian systems}
\vskip 1cm
{\it R. Mastroianni\footnote{rita.mastroianni@math.unipd.it} and C. Efthymiopoulos\\
Dipartimento di Matematica Tullio Levi-Civita, Universit\`{a} degli 
Studi di Padova}\\
via Trieste 63, 35121 Padova
\end{center}

\begin{abstract}
We present a Kolmogorov-like algorithm for the computation of a normal form in the neighborhood of an invariant torus in `isochronous' Hamiltonian systems, i.e., systems with Hamiltonians of the form $\Hscr=\Hscr_0+\varepsilon \Hscr_1$ where $\Hscr_0$ is the Hamiltonian of $N$ linear oscillators, and $\Hscr_1$ is expandable as a polynomial series in the oscillators' canonical variables. This method can be regarded as a normal form analogue of a corresponding Lindstedt method for coupled oscillators. We comment on the possible use of the Lindstedt method itself under two distinct schemes, i.e., one producing series analogous to those of the Birkhoff normal form scheme, and another, analogous to the Kolomogorov normal form scheme in which we fix in advance the frequency of the torus.
\end{abstract}

\section{Introduction and background}
In various contexts in the literature, the use of the term `Lindstedt series' for isochronous Hamiltonian systems often refers to one of two {\it distinct} methods, both applicable to the perturbative study of the dynamics around systems with elliptic equilibria. The difference between these two methods can be conveniently explained  with the help of the following example: consider a `Henon-Heiles' type Hamiltonian
\begin{equation}\label{hhnonres}
\Hscr = {1\over 2}(p_x^2+p_y^2) + {1\over 2}\omega_{0,1}^2 x^2  + {1\over 2}\omega_{0,2}^2 y^2 
+ \varepsilon P_3(x,y)
\end{equation}
where, contrary to the actual H\'{e}non-Heiles model (\cite{henhei1964}) (where $\omega_{0,1}=\omega_{0,2}=1$), we first assume that the frequencies $(\omega_{0,1},\omega_{0,2})$ satisfy no resonance condition.  $P_3$ can be any polynomial cubic in $x,y$. 

We are interested in constructing perturbative series solutions in the model (\ref{hhnonres}) under the form:
\begin{equation}\label{xyserhh}
x(t)= x_0(t)+\varepsilon x_1(t)+\varepsilon^2 x_2(t)+\ldots,~~~ 
y(t)= y_0(t)+\varepsilon y_1(t)+\varepsilon^2 y_2(t)+\ldots,~~~ 
\end{equation}
where we adopt some periodic form for the functions $x_0(t),y_0(t)$ and compute iteratively all subsequent functions $x_i(t),y_i(t)$, $i=1,2,\ldots$. We further want to secure that the iterative scheme used to compute the functions $x_i(t),y_i(t)$ preserves the quasi-periodic character of the solutions, i.e., produces no secular terms (of the form $t\sin(\omega t)$, etc) for some frequencies $\omega$ obtained as discussed below. An elementary remark in this context is that the nonlinear coupling of the oscillators implies that quasi-periodic orbits in the above model are expected to evolve, in general, with frequencies $\omega_1,\omega_2$ different from those of the unperturbed oscillators, $\omega_{0,1},\omega_{0,2}$. As it is well known (see, for example, \cite{gio1998}), recognition of this fact implies to introduce formal series also for the frequencies:
\begin{eqnarray}\label{omeserhh}
\omega_1(A_1,A_2) &= &\omega_{0,1} +\varepsilon\omega_{1,1}(A_1,A_2) +\varepsilon^2\omega_{2,1}(A_1,A_2) + \ldots,~~~\nonumber \\
\omega_2(A_1,A_2) &= &\omega_{0,2} +\varepsilon\omega_{1,2}(A_2,A_2)+  \varepsilon^2\omega_{2,2}(A_1,A_2) + \ldots.
\end{eqnarray}
The quantities $\omega_{i,1}(A_1,A_2),\omega_{i,2}(A_1,A_2)$ are functions depending on two parameters $A_1,A_2$, called hereafter the `amplitudes' of the oscillations in $x$ and $y$ respectively. They enter into the calculation through the choice made for the zero-th order terms $x_0(t)$, $y_0(t)$, since the iterative procedure starts by setting
\begin{equation}\label{xy0hh}
x_0(t) = A_1\cos(\omega_1 t+\phi_{x0}),~~~ y_0(t) = A_2\cos(\omega_2 t+\phi_{y0})~~
\end{equation}
where the initial phases $\phi_{x0},\phi_{y0}$ can be arbitrary. 

The above are common elements of the point of departure for both versions discussed below of the Lindstedt method. However, at this stage emerges an important bifurcation in the way we define the iterative scheme by which the functions $x_i(t),y_i(t),\omega_{i,1},\omega_{i,2}$ are to be computed. We discuss two distinct possibilities, referred to below as (i) a Lindstedt scheme `analogous to the Birkhoff series', or (ii) a Lindstedt scheme `analogous to the Kolmogorov series'. 

As typical in perturbation theory, the formal difference between the above two schemes actually reflects a real (physical) difference in the way we interpret the meaning of the series (\ref{omeserhh}). In summary, the difference can be posed as follows (see section \ref{sec2} for details): 

(i) in the scheme called below `analogous to Birkhoff', we seek to construct a quasi-periodic solution valid for any value of the amplitudes $A_1,A_2$ within a suitably defined open domain around the origin. Thus, the series (\ref{omeserhh}) in this scheme are meant to answer the question of what are the values of the frequencies $\omega_1,\omega_2$ under which the motion takes place for {\it any given and pre-selected} sets of values of the amplitudes $A_1,A_1$ in the above domain. The reader is referred to \cite{jormas1999} where a clear exposition of the method is given in the framework of special solutions of the three-body problem computed via Lindstedt series.  

(ii) in the scheme called below `analogous to Kolmogorov', instead, we {\it fix in advance} the values of the frequencies $\omega_1,\omega_2$ (see \cite{gio1998} for a clear exposition of the method in the context of the forced anharmonic oscillator); this is called by some authors a `torus fixing method'. A relevant remark in the context of this last method is that the series (\ref{omeserhh}) are actually purported to answer the question {\it reverse} to the one posed in (i) above. That is, the question now is: with given and pre-selected values of the frequencies $\omega_1,\omega_2$, invert the series (\ref{omeserhh}) and compute which are the corresponding amplitudes $A_1,A_2$ for which we obtain quasi-periodic trajectories with the frequencies $\omega_1,\omega_2$. Thus, in method (i) the series are parameterized by the amplitudes $A_1,A_2$, which can be selected at the beginning of the construction, while in method (ii) the solutions are parameterized by the frequencies $\omega_1,\omega_2$, which are the parameters to select at the beginning of the construction. Also, in the latter case the series inverse to (\ref{omeserhh}) turn out to have the form (in the cubic 
case)\\
\begin{equation}\label{ampserhh}
\varepsilon^2 A_1^2 = \sum_{i=1}^\infty C_{i,1} (\omega_{1}-\omega_{0,1})^i,~~~
\varepsilon^2 A_2^2 = \sum_{i=1}^\infty C_{i,2} (\omega_{2}-\omega_{0,2})^i,
\end{equation}
for some constant coefficients $C_{i,1},C_{i,2}$ computable from the series (\ref{omeserhh}). Thus, with all frequencies of the problem fixed in advance, establishing the convergence of the inverse series (\ref{ampserhh}) suffices to answer the question posed at (ii).

The question of the convergence of the series is, of course, crucial, and related to the kind, and pattern of accumulation in the series terms, of small divisors appearing at successive perturbative steps. As regards the kind of divisors, we can readily see that:

- in scheme (i) we obtain divisors of the form $k_1\omega_{0,1}+k_2\omega_{0,2}$, with $(k_1,k_2)\in\mathbb{Z}^2$, $|k_1|+|k_2|\neq 0$. This follows from the kind of linear (non-homogeneous) equation to solve iteratively. Deferring details to the example treated in Section 2, we briefly recall that in scheme (i) we introduce the parametrization (modulo two unimportant phases) $\varphi_1 = \omega_1 t$, $\varphi_2=\omega_2 t$, and after introducing the series expressions~(\ref{xyserhh}) and (\ref{omeserhh}) to the equations of motion and separate terms of  like orders we arrive at equations (to be solved iteratively) of the form:
\begin{eqnarray}\label{sch1}
&~&
\left(
 \omega_{0,1}{\partial\over\partial\varphi_1}
+\omega_{0,2}{\partial\over\partial\varphi_2}
\right)^2x_i+ x_i = \Theta_{1,i}(\varphi_1,\varphi_2) 
\\
&~&
\left(
 \omega_{0,1}{\partial\over\partial\varphi_1}
+\omega_{0,2}{\partial\over\partial\varphi_2}
\right)^2y_i+ y_i = \Theta_{2,i}(\varphi_1,\varphi_2) 
~~~i=1,2,\ldots\nonumber
\end{eqnarray}
where the functions $\Theta_{1,i}(\varphi_1,\varphi_2)$, $\Theta_{2,i}(\varphi_1,\varphi_2)$ contain trigonometric terms in the angles $\varphi_1,\varphi_2$ (see \cite{jormas1999}, section 4). 

- In scheme (ii), instead, we obtain divisors of the form $k_1\omega_1+k_2\omega_2$, i.e., depending on the (fixed) pre-selected new frequencies $\omega_1,\omega_2$. This follows from the fact that the linear non-homogeneous equations to solve are now of the form (see \cite{gio1998}):
\begin{eqnarray}\label{sch1b}
&~&
\ddot{x}_i+\omega_1^2 x_i = \Phi_{1,i}(\varphi_1,\varphi_2) 
\\
&~&
\ddot{y}_i+\omega_2^2 y_i = \Phi_{2,i}(\varphi_1,\varphi_2) 
~~~i=1,2,\ldots\nonumber
\end{eqnarray}
again with functions $\Phi_{1,i}(\varphi_1,\varphi_2)$, $\Phi_{2,i}(\varphi_1,\varphi_2)$ containing trigonometric terms in the angles $\varphi_1=\omega_1 t,\varphi_2=\omega_2 t$. Note that since the divisors depend on the new frequencies $\omega_1,\omega_2$, choosing non-resonant values for the {\it latter} permits the formal construction to proceed; this, even when the unperturbed frequencies $\omega_{0,1},\omega_{0,2}$ are, instead, resonant. 

As regards convergence, in the case (i) Poincar\'{e} (\cite{poi1893}, Ch.IX) already emphasizes that the Lindstedt series with divisors depending on the original harmonic frequencies $\omega_{0,1},\omega_{0,2}$ are divergent, exhibiting the well known asymptotic character associated with the series computed via a Birkhoff normal form (see \cite{eftetal2004} for a review). Indeed, as shown by example in section 2 below, it possible to construct Birkhoff series yielding the same individual solutions as those of the Lindstedt series of scheme (i). We note here that the series originally introduced by Lindstedt (\cite{lind1883b,lin1883a,lin1886}), albeit somewhat different in structure, exhibit the same divisors  as those of the scheme (i) above, thus, according to Poincar\'{e}, they are only asymptotic. On the other hand, Eliasson (\cite{eli1996}) and Gallavotti (\cite{gal1994a,gal1994b}) established the existence of convergent Lindstedt series by the `torus fixing method' on the basis of the cancellations between terms with small divisors (see \cite{gioloc1997} for an instructive example). A proof of the convergence of scheme (ii) is actually possible by diagrammatic methods via the following theorem \cite{coretal2011}:  

\begin{Theorem}[\cite{coretal2011}]
\label{corsietal}
Consider the $N$ coupled oscillator equations
\begin{equation}\label{oscicoup}
\ddot{x}_j + \omega^2_jx_j + f_j(x_1,\ldots,x_N;\varepsilon)+(\omega_{0,j}^2-\omega_j^2)x_j = 0,~~~~~~~
j=1,\ldots,N
\end{equation}
where $\varepsilon$ is a real parameter, $\vet{f}(\vet{x},\varepsilon)$ is real analytic at $\vet{x}=0\, ,\,\varepsilon=0$ and at least quadratic in $\vet{x}$, and the frequency vector $\vet{\omega}$ is diophantine. Let 
\begin{equation}\label{linsol0}
x^{(0)}_j(t) = c_je^{i\omega_jt}+c_j^*e^{-i\omega_jt}, ~~~
j=1,\ldots,{N}
\end{equation}
be a solution of (\ref{oscicoup}) for any choice of the complex constants $c_j$ and for $\varepsilon=0$. Let $\Gamma(\vet{c}) = \max(|c_1|,\ldots,|c_N|,1)$. Then, there exists a positive constant $\eta_0$ and a function $\vet{\eta}(\varepsilon,\vet{c})$ holomorphic in the domain $|\varepsilon|\Gamma^3(\vet{c})\leq\eta_0$, real for real $\varepsilon$, such that the system
\begin{equation}\label{oscicoup2}
\ddot{x}_j + \omega^2_jx_j + f_j(x_1,\ldots,x_{{N}},\varepsilon) +
\eta_j(\varepsilon,\vet{c})x_j= 0,~~~~~~~
j=1,\ldots,{N}
\end{equation}
admits a solution of the form
\begin{equation}\label{linsol}
\vet{x}(t,\varepsilon,\vet{c}) =  \sum_{\vet{\nu}\in\interi^{{N}}}\vet{A}_\vet{\nu}\exp(i\vet{\nu}\cdot(\vet{\omega}t))
\end{equation}
holomorphic in the domain $|\varepsilon|\Gamma^3(\vet{c})e^{3|\vet{\omega}||\Im t|}\leq \eta_0$ and real for real $\varepsilon,t$. The constants $\vet{A}_\vet{\nu}$ 
are $\mathcal{O}(\varepsilon)$, except for the constants $A_{1,0,...,0}$, 
$A_{0,1,...,0}$, $A_{0,0,...,1}$, which are equal to $c_1$, $c_2$, \ldots, 
$c_{{N}}$ respectively. 
\end{Theorem}

A similar proof in action-angle variables in the case $N=2$ is discussed in~\cite{bargen2002}.
As a final introductory remark, the series construction in the isochronous case finds a plethora of applications in various fields of physics. We mention in particular, the use of the Lindstedt method for the computation of solutions lying on low-dimensional tori (`q-tori') in the celebrated Fermi-Pasta-Ulam (FPU) problem (\cite{chretal2010,chreft2013}). The FPU model takes, in normal mode space, the form of $N$ harmonic oscillators coupled with nonlinear terms:
\begin{equation}
\label{fpu}
\ddot{Q_k}+\Omega^2_k Q = \varepsilon F_k(Q_1,\ldots,Q_N)
\end{equation}
where the frequencies $\Omega_k$, $k=1,\ldots,N$ are given in terms of the FPU normal mode spectrum $\Omega_k = 2\sin(k\pi/(2(N+1))$, the function $F$ can be cubic or quartic in the variables $Q_k$, and the perturbation $\varepsilon$ satisfies some scaling law with $N$. 

Flach and co-workers (\cite{flaetal2005,flaetal2006}) emphasized the special role for dynamics played by solutions called `q-breathers'. These are periodic orbits of the form $Q_q(t) = A_q\cos(\omega_q t+\phi_q)$, and $Q_k(t)=0$ for $k\neq q$. For the frequencies $\omega_q$ we obtain series expressions of the form 
\begin{equation}
 \label{domeq}
\omega_q = \Omega_q + \Delta\omega_q(A_q;\varepsilon),~~~~~
\Delta\omega_q = \Oscr(\varepsilon).
\end{equation}
Then, for $\varepsilon$ sufficiently small, the Lindstedt method (ii) above allows to represent the q-breathers via the Fourier expansion 
\begin{eqnarray}
\label{qbre}
Q_q(t) &=&A_q\cos(\omega_q t+\phi_q) 
+ \sum_{m=0}^\infty \tilde{f}_{q,m}(A_q;\varepsilon)\cos[m(\omega_q t +\phi_q)]\\
Q_k(t) &=&
\sum_{m=0}^\infty \tilde{f}_{k,m}(A_q;\varepsilon)\cos[m(\omega_q t +\phi_q)],~~k\neq q,
\nonumber
\end{eqnarray}
where $\tilde{f}_{k,m}=\Oscr(\varepsilon^{p(k,q,m)})$, with integer exponent $p(k,q,m)\geq 1$. The relevant point for the FPU problem is that the rules of propagation of the amplitude $A_q$ in the series terms for all modes allows to find an analytic formula explaining the phenomenon of `energy localization' observed for particular initial excitations in the FPU model. In \cite{chretal2010} and \cite{chreft2013}, on the other hand, it was shown that the q-breathers constitute only the first member in the hierarchy of special 
FPU solutions that exhibit energy localization. More general members are the `q-tori', i.e., special solutions with $M<N$ incommensurable frequencies satisfying 
\begin{equation}
 \label{domeqi}
\omega_{q_i} = \Omega_{q_i} + \Delta\omega_{q_i}(A_{q_1},A_{q_2},\ldots,A_{q_M};\varepsilon),~~~~~
\Delta\omega_{q_{{i}}} = \Oscr(\varepsilon),
\end{equation}
where $\Rscr_q=(q_1,\ldots q_M)\in \{1,2,...,N\}^M$. 
The corresponding Fourier representation of these special solutions can again been computed using Lindstedt series, and it obtains the form
\begin{eqnarray}
\label{qtor}
Q_k(t) &=& A_{k}\cos(\omega_kt+\phi_k)+\sum_{|\mathbf{m}|=0}^\infty 
\tilde{f}_{k,{\vet{m}}}(\mathbf{A};\varepsilon)\cos[\vet{m}\cdot(\vet{\omega} t 
+\vet{\phi})],~~~k\in\Rscr_q \\
Q_k(t) &= &~~~~~~~~~0~~~~~~~~~~~+\sum_{|\mathbf{m}|=0}^\infty 
\tilde{f}_{k,{\vet{m}}}(\mathbf{A};\varepsilon)\cos[\vet{m}\cdot(\vet{\omega} t 
+\vet{\phi})],~~~ k\notin\Rscr_q \nonumber
\end{eqnarray}
with $\vet{m} \equiv (m_1,\ldots,m_M)\in\mathbb{Z}^M$, $\vet{A}\equiv(A_{q_1},\ldots,A_{q_M})$, 
$\vet{\omega}\equiv(\omega_{q_1},\ldots,\omega_{q_M})$, $\vet{\phi}\equiv(\phi_{q_1},\ldots,\phi_{q_M})$, and $\tilde{f}_{k,\vet{m}}=\Oscr(\varepsilon)$ for all 
$k=1,\ldots N$.
Furthermore, the propagation of the amplitudes $A_k$ in the series terms allows to interpret a variety of complex localization profiles encountered for particular initial mode excitations in the FPU problem (see the corresponding theorems in \cite{chreft2013}). 

We mentioned already that for the Lindstedt series analogous to the Birkhoff ones there exists a Birkhoff normal form yielding the same solutions as those recovered by the Lindstedt method via an indirect approach, i.e., one based on a sequence of normalizing transformations involving canonical changes of variables. It is natural to ask whether this correspondence between a direct (Lindstedt) and indirect (normal form) method extends in the case of the torus-fixing method as well. Due to the lack of a twist condition, the torus-fixing process in the isochronous case has to be dealt with using a technique based on `counterterms' (see \cite{gal1994a}), or a KAM algorithm `with knobs' (see \cite{sandan2021} in the present volume). We have developed a Kolmogorov algorithm using counterterms, which is able to recover the solutions of the direct Lindstedt method in both cases of full or low-dimensional tori. In the present paper we will discuss in some detail the proposed Kolmogorov algorithm for the full-dimensional case only, deferring the low-dimensional case to a different publication. Section 2 gives an elementary example allowing to fix all principle ideas in this comparison. Section 3 gives the general algorithm for the Kolmogorov method in the isochronous case. This is similar as the algorithm `with knobs' discussed by Sansottera and Danesi (\cite{sandan2021}) in the present volume. We point out, however, some differences in our viewpoints regarding what the algorithm  actually achieves to compute and how it should be implemented in the context of isochronous models with given initial unperturbed frequencies $\vet{\omega}_0$. Some results and detailed proofs are deferred to the Appendix.  

\section{An elementary example}\label{sec2}
In order to illustrate the methods discussed above, we will consider an elementary example stemming from the following one-degree of freedom Hamiltonian with a even power dependence on the canonical variables
\begin{equation}
\label{Ham.es}
\Hscr(x, p)=\Hscr_0+\varepsilon \Hscr_1 = \frac{\omega_0}{2}\left(p^2+x^2\right)+\varepsilon \frac{x^4}{4}\,.
\end{equation}
Using the harmonic oscillator action-angle variables $(J,\,q)$ with $x=\sqrt{2J}\sin(q)\,$, $p=\sqrt{2 J}\cos(q)\,$, we obtain
\begin{equation}
\label{Ham.es.Jq}
\Hscr(q,J)= \omega_0 \,J+\frac{3 \,\varepsilon }{8} J^2  - \frac{\varepsilon}{2}  J^2 \cos(2 q) + \frac{\varepsilon}{8}  J^2 \cos(4 q)\, .
\end{equation}

Let $J_0$ be the label of a given torus (periodic orbit) of the harmonic oscillator model $\Hscr_0$. Consider a real neighborhood ${\cal D}_\varepsilon=\{J=J_0+p~\mbox{with}~|p|<D_\varepsilon\}$, where $D_\varepsilon={\cal O}(\varepsilon)$. We will illustrate four different perturbative methods to treat the dynamics in the phase-space neighborhood $\mathbb{T}\times{\cal D}_\varepsilon$: these are i) a Birkhoff normal form construction, with ii) its analog in terms of Lindstedt series, iii) a Lindstedt series exhibiting the torus-fixing property of the Kolmogorov method, and, finally iv) the normal form analogue of iii). In the next section, we will give the general formulas for the construction of the Kolmogorov normal form in the case of n-degree of freedom Hamiltonians with isochronous integrable part $\Hscr_0$. 

\subsection{Birkhoff normal form}
Setting $J=J_0+p\,$ the Hamiltonian takes the form (apart from a constant)
\begin{equation}
\label{Ham.es.pq}
\begin{split}
\Hscr(q,p)&=  \omega_0\, p +\frac{3\,\varepsilon\, J_0^2}{8} +\frac{3  \,J_0 \,\varepsilon}{4} p +\frac{3 \, \varepsilon}{8}p^2 -\frac{\varepsilon\,J_0^2}{2} \cos(2 q) -\varepsilon J_0 p \cos(2 q)\\
 &\phantom{=}-\frac{\varepsilon}{2} p^2 \cos(2 q)
+\frac{\varepsilon J_0^2}{8}  \cos(4 q)+\frac{\varepsilon\, J_0}{4}  p \cos(4 q)+\frac{ \varepsilon}{8} p^2 \cos(4 q)\, .
\end{split}
\end{equation}

A `Birkhoff normal form' for the Hamiltonian~\eqref{Ham.es.pq} can be computed by introducing a canonical transformation eliminating the angle $q\,$. Writing the Hamiltonian~\eqref{Ham.es.pq} as
$$ 
\Hscr(q, p)= Z_0+\varepsilon f_1\, ,
$$
where $Z_0= \omega_0 p$ and $\displaystyle{f_1=\sum_{k,l} c_{k,l}\,p^k e^{i l q}}\,$ we will define a Lie generating function $\chi^{(1)}(q,p)\,$ bringing this Hamiltonian to normal form up to terms ${\cal O}(\varepsilon)$. In the standard procedure, it is sufficient to set $\chi^{(1)}=X^{(1)}$, where 
$X^{(1)}$ solves the homological equation $L_{X^{(1)}}Z_0+\varepsilon f_1=\varepsilon \zeta_1\,$, with 
$\zeta_1=<f_1>_q$ and $L_X $ denoting the Poisson bracket operator $L_X = \{\cdot,X\}$. However, comparing the result with the one obtained by the corresponding Lindstedt method (see next subsection) requires a small modification in the definition of $\chi^{(1)}$. Consider the canonical transformation $(q,p)\rightarrow(q^{(r)},p^{(r)})$ obtained after $r$ normalization steps: we require that the canonical transformation be such that the initial condition $q^{(r)}=p^{(r)}=0$ in the new variables be mapped to the initial condition $\left(q(q^{(r)}=0,p^{(r)}=0),p(q^{(r)}=0,p^{(r)}=0)\right)=(0,0)+{\cal O}(\varepsilon^{r+1})$ in the original variables. It is easy to see that such a requirement of control on the initial condition can be fulfilled by setting 
$$
\chi^{(n)}(q,p)=X^{(n)}(q,p)+K^{(n)}\,q+S^{(n)}\,p\, ,\quad n=1,\ldots,r
$$
where $K^{(n)}$, $S^{(n)}$ are constants possible to compute at every step by requiring that 
$$
\left(q(q^{(n)}=0,p^{(n)}=0),p(q^{(n)}=0,p^{(n)}=0)\right)=(0,0)+{\cal O}(\varepsilon^{n+1}).
$$
Since $\{\omega_0 p, K^{(n)}\,q\}=-\omega_0 K^{(n)}$, this procedure will only alter the normal form at the n-th step by a constant, adding, however, some trigonometric terms to the remainder at every step. As an example, we can readily verify the following formulas for the first step (and analogously for subsequent steps) 
$$
\chi^{(1)}(q,p)=X^{(1)}(q,p)+K^{(1)}\,q+S^{(1)}\,p\, ,
$$
where 
\begin{align*}
&X^{(1)}(q,p)=\sum_{\substack{l\neq 0\\ k}} \frac{c_{k,l}}{i\,l\,\omega_0}\,p^k e^{i l q}\, , && & K^{(1)}=-\sum_{l\neq 0 } \frac{c_{0,l}}{\omega_0}\, , && &S^{(1)}=-\sum_{l\neq 0 } \frac{c_{1,l}}{i\,l\,\omega_0}\, .\end{align*} 
Omitting details, the formulas obtained after two normalization steps as above are the following: the Hamiltonian takes the form 
$$\Hscr^{(2)}(q^{(2)}, p^{(2)})=\mathrm{exp} L_{\varepsilon^2\chi^{(2)}}\mathrm{exp} L_{\varepsilon\chi^{(1)}} \Hscr\Big|_{\substack{q=q^{(2)}\\p=p^{(2)}}} = \left(Z_0+\varepsilon Z_1 +\varepsilon^2 Z_2+ \sum_{i\geq 3} \varepsilon^i f_{i}\right)\!\bigg|_{\substack{q=q^{(2)}\\p=p^{(2)}}} \, .$$
For simplicity in the notation, from now on we omit superscripts from the variables $q$ and $p$ unless explicitly required, adopting, instead, the convention that the symbols $(\q,\p)$ in any function of the form $F^{(r)}(\q,\p)$ imply the new canonical variables computed after $r$ normalization steps. Then, up to order $2$ in $\varepsilon$ we obtain:
\begin{align*}
\Hscr^{(2)}(\q, \p)=Z^{(2)}+R^{(2)}=\omega_0\,\p+\frac{3\, \varepsilon\, J_0 }{4}\,\p + \frac{3\, \varepsilon\, {\p}^2}{8} - \frac{69 \,\varepsilon^2 \,J_0^2 }{64 \,\omega_0}\,\p - \frac{
 51\, \varepsilon^2 \,J_0}{64\, \omega_0}\,{\p}^2 -\frac{17\, \varepsilon^2 }{64\, \omega_0}\,{\p}^3
 +R^{(2)}\, 
 \end{align*}
where the remainder $R^{(2)}$ is ${\cal O}(\varepsilon^3)$. The Hamiltonian $Z^{(2)}$ can now be used to analytically compute $\varepsilon^2-$precise solutions to the equations of motion in the variables $(\q,\p)$. The equations of motion are
\begin{align*}
\begin{cases}
&\displaystyle{ \dot{\q}=\frac{\partial \Hscr^{(2)}}{\partial \p}= \omega_0+\frac{3 \,\varepsilon \,J_0}{4}+\frac{3 \,\varepsilon }{4} \p-\frac{69\, \varepsilon^2 \,J_0^2}{64 \,\omega_0}-\frac{51 \,\varepsilon^2\,J_0 }{32 \,\omega_0}\,\p-\frac{51 \,\varepsilon^2}{64 \,\omega_0}\, \p^2}\\[1.7ex]
&\displaystyle{\dot{\p}=-\frac{\partial \Hscr^{(2)}}{\partial \q}=0}
\end{cases}\, ;
\end{align*}
fixing the initial condition $\p(0)=\q(0)=0\,$ yields the solution
\begin{align}\label{sol.Birk.new}
 &\p(t)=0\, , &&  &\q(t)=\omega\,t:=\varphi_B\, , && &\omega=\omega_0+\frac{3 \,\varepsilon\,J_0}{4}-\frac{69 \,\varepsilon^2 \,J_0^2}{64 \,\omega_0}\, .
\end{align} 
This can be back-transformed to the solution in the original variables. We have
\begin{align}
\label{old.new}
&q=\mathrm{exp} L_{\varepsilon^2\,\chi^{(2)}}\mathrm{exp} L_{\varepsilon\,\chi^{(1)}}\,\q\, , && &p=\mathrm{exp} L_{\varepsilon^2\,\chi^{(2)}}\mathrm{exp} L_{\varepsilon\,\chi^{(1)}}\,\p\, .
\end{align}
Substituting the solutions $(\q(t), \, \p(t))$ in the previous expression, we find
\begin{equation}\label{sol.Birk.old}
\begin{split}
q(t)&=\varphi_B-\frac{\varepsilon\, J_0 }{2 \,\omega_0}\,\sin (2 \varphi_B)+\frac{31\,\varepsilon^2\, J_0^2 }{32\, \omega_0^2}\,\sin (2 \,\varphi_B)+\frac{\varepsilon\,J_0 }{16\,\omega_0}\,\sin (4 \varphi_B)\\
&\phantom{=}-\frac{\varepsilon^2 \,J_0^2 }{32 \,\omega_0^2}\,\sin (4 \,\varphi_B)-\frac{\varepsilon^2\, J_0^2}{32 \,\omega_0^2}\, \sin (6 \,\varphi_B)+\frac{\varepsilon^2 \,J_0^2}{512 \,\omega_0^2}\, \sin (8 \,\varphi_B)\, ,\\[1.2ex]
p(t)&= - \frac{3\, \varepsilon \,J_0^2}{8 \,\omega_0}+\frac{13 \,\varepsilon^2\, J_0^3}{16\,\omega_0^2} +\frac{\varepsilon\, J_0^2 }{2\, \omega_0}\,\cos (2\,\varphi_B)-\frac{33\,\varepsilon^2\,J_0^3 }{32\,\omega_0^2}\,\cos (2\,\varphi_B)\\
&\phantom{=}-\frac{\varepsilon\,J_0^2 }{8 \omega_0}\,\cos (4 \,\varphi_B)+\frac{3 \,\varepsilon^2\,J_0^3}{16 \,\omega_0^2}\, \cos (4 \,\varphi_B)+\frac{\varepsilon^2\, J_0^3 }{32\,\omega_0^2}\,\cos (6 \,\varphi_B)\, ,\\
J(t)&=J_0+p(t)\,.
\end{split}
\end{equation} 
Observe that $q(0)=0$ and $J(0)=J_0\,$, as was required. 

\noindent Two remarks are in order:

 (i) The divisors appearing in all series expressions obtained above depend on the (unique, in the case of 1DOF systems) {\it unperturbed} frequency of the model, i.e., the frequency $\omega_0$ of the linear oscillator. In the case of systems with $N>1$ degrees of freedom, the above series will produce divisors of the form $\vet{m}\cdot\vet{\omega_0}$, where $\vet{m}\in\mathbb{Z}^N$, $|\vet{m}|\neq 0$, and $\vet{\omega_0}$ is the $N-$vector of the unperturbed frequencies of the problem. This implies that the method may formally proceed only when the vector $\vet{\omega_0}$ is non-resonant. As regards the series convergence, this is guaranteed in an open domain in the 1DOF case. However, when $N>1$ the series are in general only asymptotic (see \cite{eftetal2004} for reviews). 

 (ii)
The value of the initial datum for the action $J_0$ can be chosen at the end of the process, i.e., the numerical value of $J_0$ need not be fixed in advance. In fact, $J_0$ (or, more generally, the vector ${\vet{J}_0}\,$) can be regarded as a parameter free to transfer in all iterative computations. It is, then, 
in terms of this parameter that we obtain an expression for the frequency(ies) $\omega$ on the torus with the initial conditions $J(0)=J_0$, $q(0)=0$ as a function of $J_0$. In our example this function is given up to ${\cal O}(\varepsilon^2)$ in Eq (\ref{sol.Birk.new}). 

 (iii)
Any choice is possible to impose on the initial phase $q(0)$, altering slightly the definition of the constants $S^{(n)}$.

\subsection{Lindstedt solution analogous to Birkhoff}
Analytical solutions in the form of formal series obtained like (\ref{sol.Birk.old}) above will be hereafter called `Birkhoff solutions'. This terminology emphasizes the fact that the solutions are obtained using a Birkhoff normal form procedure. It is notworthy that Poincar\'{e}'s reference to the `Lindstedt series' (Methodes nouvelles, Ch. IX) actually consists of the construction of solutions to coupled oscillator problems obtained `indirectly', i.e., via transformations of the variables and the engineering of the corresponding  Hamiltonian function. Hence, the more general term `Poincar\'{e}-Lindstedt' series often encountered in literature. We now give, instead, the `direct' (i.e., without transformations) series construction of the Birkhoff solution (\ref{sol.Birk.old}) by implementing, instead, the original method of Lindstedt in the framework of the canonical action-angle variables of the harmonic oscillator model. 

The Hamilton's equations of motion for the Hamiltonian~\eqref{Ham.es.Jq} are
\begin{subnumcases}{}
\label{eq.moto.L.q}
&$\displaystyle{ \dot{q}(t)=\frac{\partial \Hscr}{\partial J}=
\omega_0+\frac{3 \,\varepsilon\, J}{4}-\varepsilon \,J \,\cos (2 q)+\frac{ \varepsilon \,J\,}{4} \cos (4 q)}$\\
\label{eq.moto.L.J}
&$\displaystyle{\dot{J}(t)=-\frac{\partial \Hscr}{\partial q}=
-\varepsilon\, J^2 \sin (2 q)+\frac{\varepsilon\, J^2}{2}  \sin (4 q)}\,.$
\end{subnumcases}

We perform the following steps:

{\it Step 1: time re-parametrization.} Set $\varphi=\omega\,t$ where $\omega$ is the (still unknown) frequency on the 1-torus (periodic orbit) corresponding to the solution with initial conditions $J(0)=J_0$, $q(0)=0$.  

{\it Step 2: frequency expansion.} This is the key element of the Lindstedt method. We expand $\omega$ as
\begin{equation}\label{w0_w}
\omega=\omega_0-\varepsilon\, a_1(J_0)-\varepsilon^2\,a_2(J_0)+\ldots. 
\end{equation}
Note that the corrections $a_i$ $i=1,2,\ldots$ are functions of the parameter $J_0$, i.e., of the `amplitude' (square) of the oscillations. 

{\it Step 3: expansion of the solution.} We write
\begin{equation}
\label{perturb.sol}
\begin{split}
q(\varphi)&=q_{0}(\varphi)+\varepsilon\,q_{1}(\varphi)+\varepsilon^2\,q_{2}(\varphi)+\ldots\, ,\\
J(\varphi)&=J_{0}(\varphi)+\varepsilon\,J_{1}(\varphi)+\varepsilon^2\,J_{2}(\varphi)+\ldots\, .
\end{split}
\end{equation} 
Note here a key element of the method, which is the fact that all functions $q_i,J_i$, $i=0,1,\ldots$ are considered functions of the phase $\varphi=\omega\,t\,$ rather than of the time $t$ itself. This is the key point in the differentiation of the method presented here with respect to the `torus-fixing' method presented in Subsection \ref{Lind.classico.even}. The relevant fact is that the use of the angle $\varphi$ (or, more generally, of a set of angles $\vet{\varphi}\in\mathbb{T}^N$, with $\vet{\varphi}=\vet{\omega}t$ in the N-DOF case), instead of time, allows to split the equations of motion in a sequence of linear non-homogeneous equations, whose iterative solution introduces divisors depending on the unperturbed frequencies $\vet{\omega_0}$. This is made clear in the next step.

{\it Step 4: splitting of the equations of motion in powers of $\varepsilon$ and iterative solution.} In our example, from the definition of $\varphi$ we have:
$$
 \omega\,\frac{d q(\varphi)}{d\varphi}=\frac{d q(t)}{d t}\, , \qquad\qquad
 \omega\,\frac{d J(\varphi)}{d\varphi}=\frac{d p(t)}{d t}\, .
$$
Substituting~\eqref{w0_w} and~\eqref{perturb.sol} in the equations of motion~\eqref{eq.moto.L.q} and~\eqref{eq.moto.L.J} leads to the following expressions (up to order $2$ in $\varepsilon\,$)
\begin{subequations}\label{linstbirk}
\begin{align}
&\left(\omega_0-\varepsilon\, a_1-\varepsilon^2\,a_2\right)\left(\frac{d q_{0}(\varphi)}{d\varphi}+\varepsilon\,\frac{d q_{1}(\varphi)}{d\varphi}+\varepsilon^2\,\frac{d q_{2}(\varphi)}{d\varphi}\right)\\
& = 
\omega_0 + \frac{3\,\varepsilon\, J_0(\varphi)}{4}-\varepsilon\,J_0(\varphi)\, \cos(2 \,q_0(\varphi)) +\frac{\varepsilon\,J_0(\varphi)}{4}\,\cos(4 \,q_0(\varphi))+\frac{3\,\varepsilon^2\,J_1(\varphi)}{4}-\varepsilon^2 \,J_1(\varphi)\,\cos(2\,q_0(\varphi))
\notag\\
&\phantom{=}+\frac{\varepsilon^2\,J_1(\varphi)}{4} \,\cos(4\,q_0(\varphi))+2\,\varepsilon^2\, J_0(\varphi)\,q_1(\varphi) \sin(2\,q_0(\varphi)) -\varepsilon^2 \,J_0(\varphi) \,q_1(\varphi)\,\sin(4\,\,q_0(\varphi))\notag
~~,\\[1.2ex]
&\left(\omega_0-\varepsilon\, a_1-\varepsilon^2\,a_2\right)\left(\frac{d J_{0}(\varphi)}{d\varphi}+\varepsilon\,\frac{d J_{1}(\varphi)}{d\varphi}+\varepsilon^2\,\frac{d J_{2}(\varphi)}{d\varphi}\right)\\
&=-\varepsilon \,J_0(\varphi)^2 \,\sin(2 \,q_0(\varphi))+\frac{\varepsilon\,J_0(\varphi)^2}{2}\, \sin(4\,q_0(\varphi))-2\, \varepsilon^2 \,J_0(\varphi) \,J_1(\varphi)\,\sin(2\,q_0(\varphi))\notag\\
&\phantom{=}+\varepsilon^2\, J_0(\varphi)\,J_1(\varphi)\, \sin(4\,q_0(\varphi))-2\, \varepsilon^2\, J_0(\varphi)^2\,q_1(\varphi)\, \cos(2\,q_0(\varphi))+2\, \varepsilon^2 \,J_0(\varphi)^2\,q_1(\varphi)\, \cos(4\,q_0(\varphi))\,.\notag
\end{align}
\end{subequations}
Now, in order to iteratively determine the functions $q_i(\varphi),J_i(\varphi)$, we compare the two sides of the equations of motion at equal orders. At order $0$ we have
$$
\omega_0\,\frac{d q_{0}(\varphi)}{d\varphi}=\omega_0 \, ,\qquad\omega_0\,\frac{d J_{0}(\varphi)}{d\varphi}=0\quad\Longrightarrow\quad q_0(\varphi)=\varphi\, , \qquad J_0(\varphi)=J_0~~~.
$$
Fixing the initial data as $q_0(0)=0$ and $J(0)=J_0\,$, we then arrive at $q_0(\varphi)=\varphi$ and  $J_0(\varphi)=J_0\,$. At order~1, we now have 
\begin{equation}
\begin{split}
 &\omega_0 \,\frac{d q_1(\varphi)}{d\varphi}-a_1 =\frac{3 \,J_0}{4}- J_0 \cos (2 \varphi)+\frac{J_0}{4} \cos (4 \varphi) \\
&\omega_0\, \frac{d J_1(\varphi)}{d\varphi}=-J_0^2 \sin (2 \varphi) + \frac{J_0^2}{2} \sin (4\, \varphi)\, .
\end{split}
\end{equation}
As well known, $a_1$ is determined by the requirement that no secular terms be present in the series. This yields $a_1=-3\,J_0/4\,$. Then, the Cauchy problem with the previous differential equations and the initial constants $q_1(0)=J_1(0)=0\,$ yields the solution
$$q_1(\varphi)=-\frac{J_0 }{2\, \omega_0}\,\sin (2 \varphi)+\frac{J_0 }{16\, \omega_0}\,\sin (4 \varphi)~, \qquad J_1(\varphi)=-\frac{3 \,J_0^2}{8 \,\omega_0}+\frac{J_0^2}{2\, \omega_0}\, \cos (2 \varphi)-\frac{J_0^2 }{8\,\omega_0}\,\cos (4\varphi)\, . $$ 
The process can be repeated at subsequent orders. We leave to the reader to verify the result at second order, leading eventually to
\begin{equation}\label{sol.Lind.old}
\begin{split}
q(t)&=\varphi-\frac{\varepsilon\, J_0 }{2 \,\omega_0}\,\sin (2 \varphi)+\frac{31\,\varepsilon^2\, J_0^2 }{32\, \omega_0^2}\,\sin (2 \,\varphi)+\frac{\varepsilon\,J_0 }{16\,\omega_0}\,\sin (4 \varphi)\\
&\phantom{=}-\frac{\varepsilon^2 \,J_0^2 }{32 \,\omega_0^2}\,\sin (4 \,\varphi)-\frac{\varepsilon^2\, J_0^2}{32 \,\omega_0^2}\, \sin (6 \,\varphi)+\frac{\varepsilon^2 \,J_0^2}{512 \,\omega_0^2}\, \sin (8 \,\varphi)\, ,\\[1.2ex]
J(t)&=J_0- \frac{3\, \varepsilon \,J_0^2}{8 \,\omega_0}+\frac{13 \,\varepsilon^2\, J_0^3}{16\,\omega_0^2} +\frac{\varepsilon\, J_0^2 }{2\, \omega_0}\,\cos (2\,\varphi)-\frac{33\,\varepsilon^2\,J_0^3 }{32\,\omega_0^2}\,\cos (2\,\varphi)\\
&\phantom{=}-\frac{\varepsilon\,J_0^2 }{8 \omega_0}\,\cos (4 \,\varphi)+\frac{3 \,\varepsilon^2\,J_0^3}{16 \,\omega_0^2}\, \cos (4 \,\varphi)+\frac{\varepsilon^2\, J_0^3 }{32\,\omega_0^2}\,\cos (6 \,\varphi)\, ,
\end{split}
\end{equation} 
with $\varphi=\omega\,t$ and $\displaystyle{\omega=\omega_0+\frac{3 \,\varepsilon\,J_0}{4}-\frac{69 \,\varepsilon^2 \,J_0^2}{64 \,\omega_0}}\,$. Comparing the result with Eqs~\eqref{sol.Birk.new} and~\eqref{sol.Birk.old}, we can see that the solutions $q(t)\,$, $J(t)$ of the two methods are equal.
This is easy to justify by checking the structure of the l.h.s of Eq (\ref{linstbirk}). After the 
expansion of the frequency, the differential equations to solve at subsequent steps all involve the operators $\omega_0d/d\varphi$. Hence, all divisors appearing in the series terms are in terms of the unperturbed frequency $\omega_0$. 
\subsection{Lindstedt series analogous to Kolmogorov}
\label{Lind.classico.even}
It was already pointed out that the Lindstedt series examined so far, as well as their `indirect' (normal form) counterpart, produce, in general, series which are divergent and only have an asymptotic character.
\footnote{The examples treated in this section are obvious exceptions, since, we deal, for simplicity, with 1DOF systems exhibiting no small divisors. Small divisors appear, instead, for $N>1$. }
On the other hand, Eliasson \cite{eli1996} and Gallavotti \cite{gal1994b} established the existence of {\it convergent} Lindstedt series in nonlinear Hamiltonian systems satisfying the necessary conditions for the holding of the Kolmogorov-Arnold-Moser theorem. Gallavotti (\cite{gal1994a}) presented a diagrammatic proof of the convergence of the Kolmogorov normal form also in the `twistless' case, i.e., when the size of the Hessian matrix of the unperturbed Hamiltonian $\Hscr_0(\mathbf{J})$ is not limited from below (it is equal to zero in the `isochronous' case). A constructive Kolmogorov algorithm able to deal also with the twistless case is presented, along with the demonstration of its convergence, in the present volume by Sansottera and Danesi \cite{sandan2021}). The convergence of the Lindstedt series in this case is addressed, instead, in \cite{coretal2011} (see remarks in the introduction).  

As discussed in the introduction, when series constructions analogous to Kolmogorov are sought for in the isochronous case, an important point to address is the need for performing, at the final stage of the construction, a process involving {\it series reversal}. This reversal is necessary in order to explicitely compute the solutions $q(t)$, $J(t)$ whose initial conditions $q_0,J_0$ correspond to motion on a torus with given frequency vector $\mathbf{\omega}$. The key remark is that the value of $J_0$, which parametrises the solutions, cannot be fixed in advance due to the lack of a twist condition allowing to compute the mapping ${\omega(J_0)}$. We now discuss the application of the direct (Lindstedt) method analogous to Kolmogorov in the same example as in the previous two sections, aiming to illustrate the above points. 

Fixing the frequency of the torus in the isochronous case can be implemented as described in \cite{gio1998}: assume that we target a particular solution of the equations of motion \eqref{eq.moto.L.q} and~\eqref{eq.moto.L.J} represented as a trigonometric series and evolving according to a {\it given} pre-selected frequency $\omega$. Inverting  
the expansion (\ref{w0_w}) we obtain 
\begin{equation}\label{w_w0}
\omega_0=\omega+\varepsilon\, a_1(J_0)+\varepsilon^2\,a_2(J_0)+\ldots
\end{equation}
Also, as before we expand the solution as
\begin{equation}
\begin{split}
q(t)&=q_{0}(t)+\varepsilon\,q_{1}(t)+\varepsilon^2\,q_{2}(t)+\ldots \\
J(t)&=J_{0}(t)+\varepsilon\,J_{1}(t)+\varepsilon^2\,J_{2}(t)+\ldots\,. 
\end{split}
\end{equation} 
Note, however, that this time we perform no time-reparametrization, i.e., the solutions remain expressed as functions of the time $t$. Thus, replacing the above expressions in~\eqref{eq.moto.L.q} and~\eqref{eq.moto.L.J} the equations of motion lead now to the expressions (up to order $2$ in $\varepsilon\,$)
\begin{subequations}
\begin{align}
&\dot{q_{0}}(t)+\varepsilon\,\dot{q_{1}}(t)+\varepsilon^2\,\dot{q_{2}}(t)=
 \omega+\varepsilon\,a_1+\frac{3\,\varepsilon
\,J_0(t)}{4}-\varepsilon\,J_0(t)\, \cos(2 \,q_0(t))+\frac{\varepsilon\,J_0(t)}{4}\,\cos(4 q_0(t))\\
&\phantom{\dot{q_{0}}(t)+\varepsilon\,\dot{q_{1}}(t)+\varepsilon^2\,\dot{q_{2}}(t)=} +\frac{3\,\varepsilon
^2\,J_1(t)}{4}+\varepsilon^2 \,a_2-\varepsilon^2\,J_1(t)\,\cos(2\,q_0(t)) +\frac{\varepsilon^2\,J_1(t)}{4}\,\cos(4 \,q_0(t))\notag\\[1ex]
&\phantom{\dot{q_{0}}(t)+\varepsilon\,\dot{q_{1}}(t)+\varepsilon^2\,\dot{q_{2}}(t)=} +2\,\varepsilon^2\,J_0(t)\,q_1(t)\, \sin(2\, q_0(t))-\varepsilon^2\,J_0(t)\,q_1(t)\,\sin(4 \,q_0(t))\, , 
\notag\\[1.2ex]
&\dot{J_{0}}(t)+\varepsilon\,\dot{J_{1}}(t)+\varepsilon^2\,\dot{J_{2}}(t)=-\varepsilon \,J_0(t)^2 \,\sin (2 \,q_0(t))+\frac{\varepsilon\,J_0(t)^2}{2} \, \sin (4 q_0(t))\\
&\phantom{\dot{J_{0}}(t)+\varepsilon\,\dot{J_{1}}(t)+\varepsilon^2\,\dot{J_{2}}(t)=} -2 \,\varepsilon^2\,J_0(t)\,J_1(t) \,\sin (2 \,q_0(t))+\varepsilon^2\,J_0(t)\,J_1(t)\, \sin (4q_0(t))\notag\\[1ex]
&\phantom{\dot{J_{0}}(t)+\varepsilon\,\dot{J_{1}}(t)+\varepsilon^2\,\dot{J_{2}}(t)=}-2 \,\varepsilon^2 \,J_0(t)^2 \,q_1(t) \,\cos (2 q_0(t))+2 \,\varepsilon^2 \,J_0(t)^2\,q_1(t) \,\cos (4 \,q_0(t))\, .\notag
\end{align}
\end{subequations}
Collecting now the terms of equal order we can compute iteratively all the functions $q_i(t),J_i(t)$, for $i=0,1,2$, setting, at order zero:
$$
q_0(t)=\omega\,t\,,\qquad J_0(t)=J_0 
$$
which corresponds to the choice of the initial condition $q_0(0)=0$ and $J_0(0)=J_0\,$. Then, at first order we have
\begin{equation}\label{lindord1}
\begin{split}
&\displaystyle{ \dot{q_{1}}(t)=
 a_1+\frac{3 \, J_0}{4}-J_0 \,\cos (2 \,\omega\,t)+\frac{  J_0\,}{4} \cos (4 \,\omega\,t)}\, ,\\[1.7ex]
&\displaystyle{\dot{J_{1}}(t)=
- J_0^2 \sin (2 \,\omega\,t)+\frac{ J_0^2}{2}  \sin (4 \,\omega\,t)}
\, 
\end{split}
\end{equation}
implying $a_1=-3 \, J_0/4\,$ and 
$$
q_1(t)=-\frac{J_0 }{2\, \omega}\,\sin (2 \,\omega\,t)+\frac{J_0}{16 \,\omega}\, \sin (4 \,\omega\,t)\, ,\qquad J_1(t)=-\frac{3\,J_0^2}{8 \,\omega}+\frac{J_0^2 }{2\, \omega}\,\cos (2\,\omega\,t)-\frac{J_0^2 }{8\, \omega}\,\cos (4 \,\omega\,t).
$$
Note that the integration constants in (\ref{lindord1}) were set as $q_1(0)=J_1(0)=0\,$, consistent with our choice of initial condition. Repeating the procedure at second order yields
\begin{equation}
\begin{split}
\label{sol.Lind_classic.old}
q(t)&=\omega\,t -\frac{\varepsilon \,J_0 }{2 \,\omega}\,\sin (2\,\omega\,t)+\frac{\varepsilon \,J_0}{16 \,\omega}\, \sin (4\,\omega\,t)+\frac{19 \,\varepsilon^2 J_0^2 }{32\, \omega^2}\,\sin (2\, \omega\,t)+\frac{\varepsilon^2\,J_0^2 }{64 \,\omega^2}\,\sin (4 \,\omega\,t)\\
&\phantom{=}-\frac{\varepsilon^2 \,J_0^2 }{32 \,\omega^2}\,\sin (6 \,\omega\,t)+\frac{\varepsilon^2\, J_0^2 }{512 \,\omega^2}\, \sin (8\, \omega\,t)\, ,\\[1.7ex]
J(t)&=J_0-\frac{3\, \varepsilon\, J_0^2}{8 \,\omega}+\frac{\varepsilon\, J_0^2 }{2 \,\omega}\,\cos (2 \,\omega\,t)-\frac{\varepsilon\, J_0^2 }{8 \,\omega}\,\cos (4 \,\omega\,t)+\frac{17 \,\varepsilon^2\, J_0^3}{32 \,\omega^2}\\
&\phantom{=}-\frac{21 \,\varepsilon^2 \,J_0^3 }{32 \,\omega^2}\,\cos (2 \,\omega\,t)+\frac{3 \,\varepsilon^2 \,J_0^3 }{32 \,\omega^2}\,\cos (4 \,\omega\,t)+\frac{\varepsilon^2 \,J_0^3 }{32 \,\omega^2}\,\cos (6\,\omega\,t)\, ,\\[1.2ex]
\omega&=\omega_0+\frac{3 \,\varepsilon\,J_0}{4}-\frac{69 \,\varepsilon^2\,J_0^2}{64 \,\omega}\, .
\end{split}
\end{equation} 
It is instructive to compare the solutions (\ref{sol.Lind_classic.old}) above with those (Eq (\ref{sol.Lind.old})) found in the case of the original Lindstedt method. We have the following remarks:

(i) From second order and beyond the coefficients in front of the harmonics $\cos(m\omega t),\sin(m\omega t)$ for the same $m$ are not all equal in the two solutions. 

(ii) Most importantly, the structure of the divisors in the two solutions is different. In the torus fixing case (Eq (\ref{sol.Lind_classic.old})), all divisors involve the {\it corrected frequency}, $\omega$, instead of the original frequency $\omega_0$ of the linear oscillator. This is not a problem for the method to proceed, since this frequency is known in advance. In $N-$DOF systems, in general, with the `torus fixing method' we obtain divisors of the form $\vet{m}\cdot\vet{\omega}$. Thus, the method can proceed even when the original frequencies of the $N$ linear oscillators $\vet{\omega_0}$ are resonant, as long as we impose a non-resonant detuning in the adopted values for the final frequencies $\vet{\omega}$ on the torus.  

(iii) On the other hand, the value of the initial condition $J_0$ leading to motion on the torus with frequency $\omega$ remains uknown up to the end of the construction. From the point of view of the symbolic implementation of the method in the computer, $J_0$ is a symbol whole powers have to be carried on along with the remaining powers of trigonometric monomials in all series terms and at all iterative steps.

(iv) At the end of the process, however, $J_0$ can be estimated by reversing the series (\ref{w_w0}), given that the functions $a_n(J_0)$ are monomials of degree $n$ in $J_0$. In an analogous way, in the $N-$DOF case we will end up with $N$ series equations of the form
\begin{equation}\label{ndofj0}
\omega_{0,j}-\omega_j = \sum_{n=1}^{\infty}\varepsilon^n a_{n,j}(J_{0,1},\ldots,J_{0,N}) 
\end{equation}
where the functions $a_{n,j}(J_{0,1},\ldots,J_{0,N})$ have all been specified iteratively up to a maximum truncation order in $n$, and they are polynomial in $\vet{J_0}\equiv(J_{0,1},\ldots,J_{0,N})$. Thus, the series (\ref{ndofj0}) can be formally inverted, yielding
\begin{equation}\label{ndofj0inv}
J_{0,j}= {1\over\varepsilon}\sum_{n=1}^{\infty}P_{n,j}(\vet{\omega}-\vet{\omega_0}) 
\end{equation}
where the functions $P_{n,j}(\vet{\omega}-\vet{\omega_0})$ are polynomial of degree $n$ in the differences $(\vet{\omega}-\vet{\omega_0})$. Note that for the inverse series to converge we must require the difference $|(\vet{\omega}-\vet{\omega_0})|$ to be smaller than $\varepsilon$, a fact which limits how far we can detune $\vet{\omega}$ from $\vet{\omega_0}$ in order to be able to specify the corresponding initial condition $\vet{J_0}$. An obvious choice is $|(\vet{\omega}-\vet{\omega_0})|=\mathcal{O}(\varepsilon^2)$. At any rate, we emphasize that the convergence of the inverse series (\ref{ndofj0}) is an open problem, crucial to the applications.\footnote{In the lack of a known answer to this problem, we can adopt two different attitudes: we may fix $\vet{J_0}$ and claim that our series answer the question of what was the value of $\vet{\omega_0}$ in the original Hamiltonian corresponding to the motion on the particular torus with frequency $\vet{\omega}$ (see \cite{sandan2021}). On the other hand, in real world applications the only certainty we have is about our initial model (!), i.e., $\vet{\omega_0}$, so, we must devise a method to propose values of $\vet{\omega}$ for which the method will probably turn to converge. In the case of the FPU problem, a numerical procedure to move within the space of the parameters $\vet{\omega}$ so at to choose plausible values was proposed in \cite{chreft2013}. } 

\subsection{Kolmogorov normal form}
\label{Kam.even}
We finally arrive at the here proposed Kolmogorov normal form algorithm yielding solutions equivalent to those discussed in the previous subsection. This is implemented by the following steps: 

\noindent{\it Step 1: substitution of the frequency series into the Hamiltonian.} 
In our example, we substitute $\omega_0$ in the Hamiltonian~\eqref{Ham.es.pq} with the series~\eqref{w_w0}. This yields (up to second order, apart from constants)
\begin{align}
\label{Ham.Kam0}
\begin{split}
\Hscr(q,p)&= (\omega+\varepsilon a_1+\varepsilon^2 a_2+\ldots)\, p +\frac{3\,\varepsilon\, J_0^2}{8} +\frac{3  \,J_0 \,\varepsilon}{4} p +\frac{3 \, \varepsilon}{8}p^2 -\frac{\varepsilon\,J_0^2}{2} \cos(2 q) -\varepsilon J_0 p \cos(2 q)\\
 &\phantom{=}-\frac{\varepsilon}{2} p^2 \cos(2 q)
+\frac{\varepsilon J_0^2}{8}  \cos(4 q)+\frac{\varepsilon\, J_0}{4}  p \cos(4 q)+\frac{ \varepsilon}{8} p^2 \cos(4 q)+\ldots\, .
\end{split}
\end{align}

\noindent{\it Step 2: normalization.}
To set the Hamiltonian into Kolmogorov normal form up to second order, we fix the value of the constants $a_1,a_2$ and we perform a sequence of Lie transformations aiming to give the Hamiltonian the form (in the transformed variables) $\Hscr^{(2)}(q,p)=Z^{(2)}(q,p)+R^{(2)}(q,p)$ where the remainder $R^{(2)}$ is $\mathcal{O}(\varepsilon^3)$, while $Z^{(2)}(q,p)$ has the form $Z^{(2)}(q,p)=\omega\,p+ \varepsilon\,R_1(q,p)+\varepsilon^2\,R_2(q,p)$ with both $R_i(q,p)=\Oscr(\norm{p}^2)$ $i=1,\,2$. To this end:
\\ - First order: we fix $a_1$ so that the linear term $\varepsilon a_1 p$ acts as counterterm for the term $\varepsilon (3J_0/4)p$. This provides, in the twistless case, a process by which the frequency $\omega$ can be kept fixed (in the usual twist case, instead, this would have been accomplished by exploiting the Hessian matrix of $\Hscr_0\,$). Formally, we require that 
\begin{align}
\label{a1.KAM}
\avg{ h_{1,1}^{(0)}}_{{q}}=\avg{a_1\, p-J_0\, p \,\cos (2 q)+\frac{J_0}{4} \, p \cos (4 q)+\frac{3 \,J_0\, p}{4}}_{{q}}=0~~, 
\end{align}
leading to $a_1=-\frac{3\,J_0}{4}$. Now, we insert this expression for $a_1$ in the Hamiltonian~\eqref{Ham.Kam0}, leading to
\begin{equation}
\begin{split}
\Hscr(q,p)&= \omega\, p +\frac{3\,\varepsilon\, J_0^2}{8}  +\frac{3 \, \varepsilon}{8}p^2 -\frac{\varepsilon\,J_0^2}{2} \cos(2 q) -\varepsilon J_0 p \cos(2 q)-\frac{\varepsilon}{2} p^2 \cos(2 q)\\
 &\phantom{=}
+\frac{\varepsilon J_0^2}{8}  \cos(4 q)+\frac{\varepsilon\, J_0}{4}  p \cos(4 q)+\frac{ \varepsilon}{8} p^2 \cos(4 q)+\varepsilon^2\,a_2\,p\, .
\end{split}
\end{equation}
We can now eliminate the $O(\varepsilon)$ trigonometric terms in the Hamiltonian with the usual procedure. Namely, we define a generating function $X^{(1)}(q)\,$ used to eliminate terms not depending on the action $p\,$. These are 
$$
h_{1,0}^{(0)}=\frac{3\,J_0^2}{8}-\frac{J_0^2}{2} \cos(2 q)+\frac{J_0^2}{8} \cos (4 q)\, ,
$$
leading to $L_{X^{(1)}(q)}\left(\omega\,p\right) + h_{1,0}^{(0)}=\avg{h_{1,0}^{(0)}}_{{q}}\,$, that is 
$$
X^{(1)}(q)=\frac{J_0^2 }{32 \,\omega}\,\sin (4 q)-\frac{J_0^2 }{4\, \omega}\,\sin (2 q)\,.
$$
Note that, similarly as in the Birkhoff case (Subsection 2.1), here too we have to fix the initial conditions so that the relation $q(0)=p(0)=0$ is preserved between variables before and after the canonical transformation. This is achieved by setting the generating function as $\chi_{1}^{(1)}(q)=X^{(1)}(q) + K^{(1)}q\,$, where $K^{(1)}=3 \,J_0^2/(8\, \omega)$ is a constant. The general rules for the determination of the constants $K^{(j)}$ will be discussed in Section~\ref{KAM}. 

Using the generating function $\chi_{1}^{(1)}(q)$ we obtain the intermediate Hamiltonian
$$
\widehat{\Hscr}^{(1)}=\exp\left(L_{\varepsilon\,\chi_1^{(1)}(q)}\right)\Hscr\, ,
$$
given by
\begin{equation}
\begin{split}
\widehat{\Hscr}^{(1)}&=\omega\,p +\frac{3 \,\varepsilon}{8}\, p^2-\varepsilon\,J_0 \,p \cos (2 q)-\frac{\varepsilon}{2} \, p^2 \cos (2 q)+\frac{\varepsilon\,J_0}{4} \, p \cos (4 q)+\frac{\varepsilon}{8}\, p^2 \cos (4 q)\\
&\phantom{=} -\frac{17 \,\varepsilon^2 \,J_0^3}{64 \,\omega}+\varepsilon^2\,a_2\, p-\frac{35 \,\varepsilon^2 \,J_0^2 }{64 \omega}\,p +\frac{\varepsilon^2\,J_0^3 }{2 \,\omega}\,\cos (2 q)+\frac{7 \,\varepsilon^2\,J_0^2 \,p}{8 \,\omega}\, \cos (2 q)-\frac{11 \,\varepsilon^2\,J_0^3}{32 \,\omega}\, \cos (4 q)\\
&\phantom{=}-\frac{7 \,\varepsilon^2\,J_0^2\, p}{16 \,\omega}\, \cos (4 q)+\frac{\varepsilon^2\, J_0^3 }{8 \,\omega}\,\cos (6 q)+\frac{\varepsilon^2\,J_0^2 \,p}{8 \,\omega}\, \cos (6 q)-\frac{\varepsilon^2\, J_0^3 }{64 \,\omega}\,\cos (8 q) -\frac{\varepsilon^2 \,J_0^2\, p}{64 \,\omega}\, \cos (8 q)\, .
\end{split}
\end{equation}
We will now eliminate the trigonometric terms linear in the momentum in $\widehat{\Hscr}^{(1)}$. These are
$$
\widehat{h}_{1,1}^{(1)}=- J_0\, p \cos (2 q)+\frac{1}{4} \,J_0\,p \cos (4 q)\, 
$$
and can be eliminated by a generating function of the form $\chi_{2}^{(1)}(q,p)=\tilde{\chi}_{2}^{(1)}(q,p) + S^{(1)}\,p$ satisfying the homological equation $L_{\chi_{2}^{(1)}(q,p)}\left(\omega\,p\right)+\widehat{h}_{1,1}^{(1)}=0\,$. As before, the value of the constant $S^{(1)}$ is fixed by the requirement that the initial phase of the solution of $q$ be preserved to zero by the transformation. We find  $S^{(1)}=0\,$ (all constants $S^{(i)}=0$ when the Hamiltonian has an even symmetry). Then 
$$
\tilde{\chi}_{2}^{(1)}(q,p)=-\frac{J_0\, p }{2 \,\omega}\,\sin (2 q)+\frac{J_0 \,p}{16 \,\omega}\, \sin (4 q)\, .$$
The new Hamiltonian is
$$
\Hscr^{(1)}=\exp\left(L_{\varepsilon\,\chi_{2}^{(1)}(q,p)}\right)\widehat{\Hscr}^{(1)}\, $$
and it is in Kolmogorov normal form up to order $\varepsilon\,$. 

Repeating the same procedure at second order we arrive at the formulas
\begin{align}
\label{a2.KAM}
a_2&=\frac{69 \,J_0^2}{64 \,\omega}\, ,\\
\chi_{1}^{(2)}&=-\frac{17 \, J_0^3\, q}{64 \,\omega^2}+\frac{ J_0^3 }{4\, \omega^2}\,\sin (2 q)-\frac{11 \, J_0^3 }{128\, \omega^2}\,\sin (4 q)+\frac{ J_0^3}{48\,\omega^2}\, \sin (6 q)-\frac{ J_0^3}{512 \,\omega^2}\, \sin (8 q)\, ,\notag\\
\chi_{2}^{(2)}&=\frac{37 \,J_0^2\, p}{64 \,\omega^2}\, \sin (2 q)-\frac{7\, J_0^2\, p}{64 \,\omega^2}\, \sin (4 q)+\frac{ J_0^2 \,p}{64 \,\omega^2}\, \sin (6 q)-\frac{J_0^2\, p}{512\,\omega^2}\, \sin (8 q)\, ,\notag
\end{align}
leading to the Hamiltonian $\Hscr^{(2)}(\q, \p)=Z^{(2)}(\q, \p)+R^{(2)}(\q, \p)$
with the Kolmogorov normal form part
\begin{align*}
Z^{(2)}(\q, \p)&=\omega\,\p  +\frac{3 \,\varepsilon\, \p^2}{8}-\frac{\varepsilon\,\p^2}{2} \, \cos (2 \q)+\frac{\varepsilon\,\p^2}{8} \, \cos (4 \q) -\frac{51\,\varepsilon^2\,J_0\, \p^2}{64 \,\omega}+\frac{37 \,\varepsilon^2\,J_0\, \p^2}{32 \,\omega}\, \cos (2 \q)\\
&\phantom{=}-\frac{7 \,\varepsilon^2\,J_0 \,\p^2}{16 \,\omega}\, \cos (4 \q)+\frac{3 \,\varepsilon^2\,J_0\, \p^2}{32 \,\omega}\, \cos (6 \q)-\frac{\varepsilon^2 \,J_0\, \p^2}{64 \,\omega}\, \cos (8 \q)\, 
\end{align*}
($(\q, \p)$ denote again the new variables after new normalization steps).

\noindent{\it Step 3: calculation of the solution on the torus.} 
Using the compact notation 
$$
Z^{(2)}(\q, \p)=\omega\,\p+\varepsilon\,R_{1}(\q,\p) + \varepsilon^2\,R_{2}(\q,\p)\, ,
$$ 
where $R_j(\q,\p)=\Oscr(\norm{\p}^2)$ $j=1,\,2\,$,  the equations of motions under the Hamiltonian $Z^{(2)}$ are
$$
\begin{cases}
&\displaystyle{ \dot{\q}=\frac{\partial Z^{(2)}}{\partial \p}=\omega+\varepsilon\,\frac{\partial R_1(\q,\p)}{\partial \p}+\varepsilon^2\,\frac{\partial R_2(\q,\p)}{\partial \p}=\omega+\Oscr(\norm{\p})}\\
&\displaystyle{\dot{\p}=-\frac{\partial Z^{(2)}}{\partial \q}=-\varepsilon\,\frac{\partial R_1(\q,\p)}{\partial \q}-\varepsilon^2\,\frac{\partial R_2(\q,\p)}{\partial \q}=\Oscr(\norm{\p}^2)}
\end{cases}\, .
$$
The torus $\p(t)=0\,$, $\q(t)=\omega\,t$ (where we chose $\q(0)=0\,$) is a solution of this system. This can be back-transformed in the original variables using 
\begin{align}
&q=\mathrm{exp} L_{\varepsilon^2\,\chi_2^{(2)}}\mathrm{exp} L_{\varepsilon\,\chi_2^{(1)}}\,\q\, , && &p=\mathrm{exp} L_{\varepsilon^2\,\chi_2^{(2)}}\mathrm{exp} L_{\varepsilon^2\,\chi_1^{(2)}}\mathrm{exp} L_{\varepsilon\,\chi_2^{(1)}}\mathrm{exp} L_{\varepsilon\,\chi_1^{(1)}}\,\p\, .
\end{align}
Substituting the solution $(\q(t)=\omega\,t, \, \p(t)=0)$ in the previous expressions, we readily deduce the solution in the original variables:
\begin{equation}
\begin{split}
\label{sol.KAM}
q(t)&=\omega\,t -\frac{\varepsilon \,J_0 }{2 \,\omega}\,\sin (2\,\omega\,t)+\frac{\varepsilon \,J_0}{16 \,\omega}\, \sin (4\,\omega\,t)+\frac{19 \,\varepsilon^2 J_0^2 }{32\, \omega^2}\,\sin (2\, \omega\,t)+\frac{\varepsilon^2\,J_0^2 }{64 \,\omega^2}\,\sin (4 \,\omega\,t)\\
&\phantom{=}-\frac{\varepsilon^2 \,J_0^2 }{32 \,\omega^2}\,\sin (6 \,\omega\,t)+\frac{\varepsilon^2\, J_0^2 }{512 \,\omega^2}\, \sin (8\, \omega\,t)\, ,\\[1.7ex]
p(t)&=-\frac{3\, \varepsilon\, J_0^2}{8 \,\omega}+\frac{\varepsilon\, J_0^2 }{2 \,\omega}\,\cos (2 \,\omega\,t)-\frac{\varepsilon\, J_0^2 }{8 \,\omega}\,\cos (4 \,\omega\,t)+\frac{17 \,\varepsilon^2\, J_0^3}{32 \,\omega^2}\\
&\phantom{=}-\frac{21 \,\varepsilon^2 \,J_0^3 }{32 \,\omega^2}\,\cos (2 \,\omega\,t)+\frac{3 \,\varepsilon^2 \,J_0^3 }{32 \,\omega^2}\,\cos (4 \,\omega\,t)+\frac{\varepsilon^2 \,J_0^3 }{32 \,\omega^2}\,\cos (6\,\omega\,t)\, ,\\
J(t)&=J_0+p(t)\, .
\end{split}
\end{equation} 
Also, using the computed expressions of $a_1$ and $a_2$ (Eqs~\eqref{a1.KAM} and~\eqref{a2.KAM}), we obtain the relation between the torus frequency and the original frequency 
$$
\omega=\omega_0+\frac{3 \,\varepsilon\,J_0}{4}-\frac{69 \,\varepsilon^2\,J_0^2}{64 \,\omega}\,  .
$$ 
These expressions are identical to the ones found by the Lindstedt method of Subsection 2.3 (see~\eqref{sol.Lind_classic.old}).

\subsection{Comparisons and numerical tests}

In order to better visualize the differences between the methods (i) and (ii) (i.e., respectively, `analogous to Birkhoff' and `analogous to Kolmogorov') we report, in the Tables~\ref{tab:q} and~\ref{tab:J} below, the series terms corresponding to the solutions for $q(t)$ and $J(t)$ as obtained by the two methods up to order $\varepsilon^2\,$.

\begin{table}[H]
	\caption{Comparison between the series terms for the solution $q(t)\,$ in the Lindstedt series obtained by the methods (i) and (ii) up to order $\varepsilon^2\,$.}
	\label{tab:q}
\resizebox{1\textwidth}{!}{
\begin{centering}
\begin{tabular}{llll}
\hline 
\rule[-4mm]{0mm}{1.cm}
Method: & $\Oscr(1)$ & $\Oscr(\varepsilon)$ & $\Oscr(\varepsilon^2) $  \\
\hline 
\rule[-1.3cm]{0mm}{2.7cm}
\makecell{(i) `Analogous\\to Birkhoff'} & $\displaystyle{\omega\,t}$ & $\displaystyle{ -\frac{ J_0 }{2 \,\omega_0}\,\sin (2\, \omega\,t)+\frac{J_0 }{16\,\omega_0}\,\sin (4\,\omega\,t)}$ & $\displaystyle{\frac{31\, J_0^2 }{32\, \omega_0^2}\,\sin (2 \,\omega\,t)-\frac{J_0^2 }{32 \,\omega_0^2}\,\sin (4 \,\omega\,t)\quad}\atop{\displaystyle{-\frac{ J_0^2}{32 \,\omega_0^2}\, \sin (6 \,\omega\,t)+\frac{J_0^2}{512 \,\omega_0^2}\, \sin (8 \,\omega\,t)}} $ \\
\rule[-1.3cm]{0mm}{2.7cm}
\makecell{(ii) `Analogous\\to Kolmogorov'} & $\displaystyle{\omega\,t}$ & $\displaystyle{-\frac{J_0 }{2 \,\omega}\,\sin (2\,\omega\,t)+\frac{J_0}{16 \,\omega}\, \sin (4\,\omega\,t)}$ & $ \displaystyle{\frac{19 \, J_0^2 }{32\, \omega^2}\,\sin (2\, \omega\,t)+\frac{J_0^2 }{64 \,\omega^2}\,\sin (4 \,\omega\,t)\quad}\atop{\displaystyle{-\frac{J_0^2 }{32 \,\omega^2}\,\sin (6 \,\omega\,t)+\frac{ J_0^2 }{512 \,\omega^2}\, \sin (8\, \omega\,t)}}$ \\
\hline 
\end{tabular}
\end{centering}
}
\end{table}

\begin{table}[H]
	\caption{Comparison between the series terms for the solution $J(t)\,$ in the Lindstedt series obtained by the methods (i) and (ii) up to order $\varepsilon^2\,$.}
	\label{tab:J}
\resizebox{1\textwidth}{!}{
\begin{centering}
\begin{tabular}{llll}
\hline 
\rule[-4mm]{0mm}{1.cm}
Method: & $\Oscr(1)$ & $\Oscr(\varepsilon)$ & $\Oscr(\varepsilon^2) $  \\
\hline 
\rule[-1.3cm]{0mm}{2.7cm}
\makecell{(i) `Analogous\\to Birkhoff'} & $\displaystyle{J_0}$ & $\displaystyle{ - \frac{3\, J_0^2}{8 \,\omega_0}+\frac{J_0^2 }{2\, \omega_0}\,\cos (2\,\omega\,t)-\frac{J_0^2 }{8 \omega_0}\,\cos (4 \,\omega\,t)}$ & $\displaystyle{\frac{13 \, J_0^3}{16\,\omega_0^2} -\frac{33\,J_0^3 }{32\,\omega_0^2}\,\cos (2\,\omega\,t)\qquad\qquad\quad}\atop{\displaystyle{+\frac{3 \,J_0^3}{16 \,\omega_0^2}\, \cos (4 \,\omega\,t)+\frac{ J_0^3 }{32\,\omega_0^2}\,\cos (6 \,\omega\,t)}} $ \\
\rule[-1.3cm]{0mm}{2.7cm}
\makecell{(ii)`Analogous\\to Kolmogorov'} & $\displaystyle{J_0}$ & $\displaystyle{-\frac{3\, J_0^2}{8 \,\omega}+\frac{ J_0^2 }{2 \,\omega}\,\cos (2 \,\omega\,t)-\frac{ J_0^2 }{8 \,\omega}\,\cos (4 \,\omega\,t)}$ & $ \displaystyle{\frac{17 \, J_0^3}{32 \,\omega^2}-\frac{21 \,J_0^3 }{32 \,\omega^2}\,\cos (2 \,\omega\,t)\qquad\qquad\quad}\atop{\displaystyle{+\frac{3  \,J_0^3 }{32 \,\omega^2}\,\cos (4 \,\omega\,t)+\frac{J_0^3 }{32 \,\omega^2}\,\cos (6\,\omega\,t)}}$ \\
\hline 
\end{tabular}
\end{centering}
}
\end{table}

From Table~\ref{tab:q} above, we observe that the solutions $q(t)$ and $J(t)$ obtained by methods (i) and (ii) have the same form up to order $\Oscr(\varepsilon)\,$, except for the fact that in method (i) all divisors depend on the frequency $\omega_0$ rather than $\omega\,$. On the other hand, the coefficients found by the two methods start differing from the order $\varepsilon^2$ and beyond. Moreover, the series for the frequency $\omega$ obtained by the two methods are:
\begin{align*}
&\omega=\omega_0+\frac{3 \,\varepsilon\,J_0}{4}-\frac{69 \,\varepsilon^2 \,J_0^2}{64 \,\omega_0}\, , && &\omega=\omega_0+\frac{3 \,\varepsilon\,J_0}{4}-\frac{69 \,\varepsilon^2\,J_0^2}{64 \,\omega}\, ,
\end{align*}
which differ again as regards their divisors.

The effect of these differences on the precision of the two methods can be seen even in the simplest case of a system with one degree of freedom in which both series are convergent (the series `analogous to Birkhoff' are, instead, divergent when $N\geq2\,$). To this end, we report below a comparison between the numerical solutions and the analytical ones, obtained by the methods (i) and (ii) in the example of the Hamiltonian~\eqref{Ham.es}. To produce the Lindstedt series (ii) in this example we work as follows: fixing the initial and final frequency $\omega_0$ and $\omega\,$, we reverse the series~\eqref{w0_w} according to
\begin{equation}
\label{inv_series}
d\omega=-\sum_{i\geq 0}\varepsilon^i\,a_i(J_0):=f(J_0)\quad\Longrightarrow\quad J_0=f^{-1}(d\omega)\,,
\end{equation}
where $d\omega=\omega-\omega_0\,$ and $f^{-1}$ denotes the series inverse to $f$. Then, having specified $J_0$ through the inverse series~\eqref{inv_series}, we compute all numerical coefficients in the Lindstedt series (i) and (ii) up to order $\varepsilon^4\,$. We analyze three different cases:
\begin{enumerate}
\item [1)]  $\varepsilon=1\,$, $\omega_0=1$ and $\omega=1.002\,$;
\item [2)] $\varepsilon=1\,$, $\omega_0=1$ and $\omega=1.02\,$;
\item [3)] $\varepsilon=1\,$, $\omega_0=1$ and $\omega=1.2\,$.
\end{enumerate}

Let us note that all results are rescalable to different choices of $\varepsilon\,$.
Reversing the series~\eqref{w0_w} as prescribed by~\eqref{inv_series}, we obtain, respectively, the following values for the amplitude (initial action datum): $J_0= 0.0026769 \,,$ $0.0277048\,, 0.383509\,$. Substituting these values of $\varepsilon\,$, $\omega\,$, $\omega_0$ and $J_0$ in the solutions $q(t)$ and $J(t)$ we obtain the solutions, as functions of $t\,$, produced by methods (i) and (ii). At the same time, it is possible to integrate the equations of motion~\eqref{eq.moto.L.q} and~\eqref{eq.moto.L.J} to produce a numerical solution, starting from the initial conditions $q(0)=0\,$, $J(0)=J_0\,$.

\begin{figure}[h!]

\frame{
\subfloat{
\begin{tikzpicture}
\node[] at (0,0)
{\begin{minipage}[b]{.3\textwidth}
\centering
\includegraphics[scale=0.193]{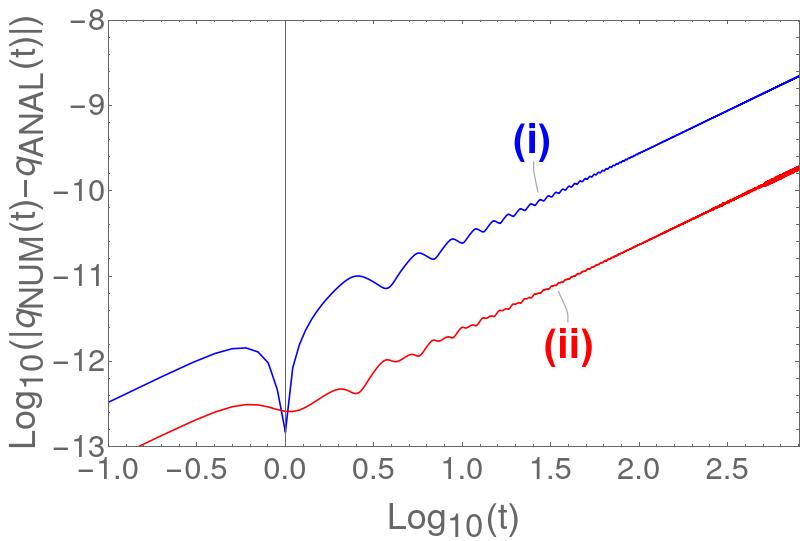}
\end{minipage}};
\end{tikzpicture}
}
\subfloat{
\begin{tikzpicture}
\node[] at (0,0)
{
\begin{minipage}[b]{.3\textwidth}
\centering
\includegraphics[scale=0.193]{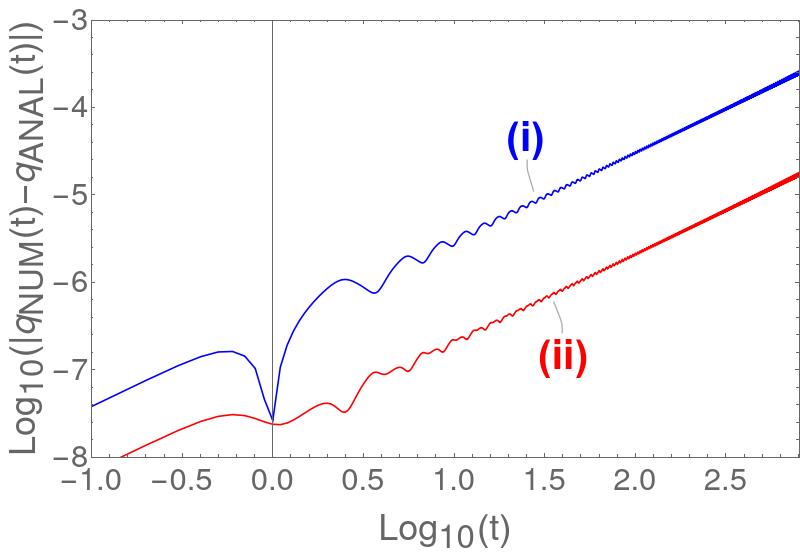}
\end{minipage}};
\end{tikzpicture}
}
\subfloat{
\begin{tikzpicture}
\node[] at (0,0)
{
\begin{minipage}[b]{.3\textwidth}
\centering
\includegraphics[scale=0.193]{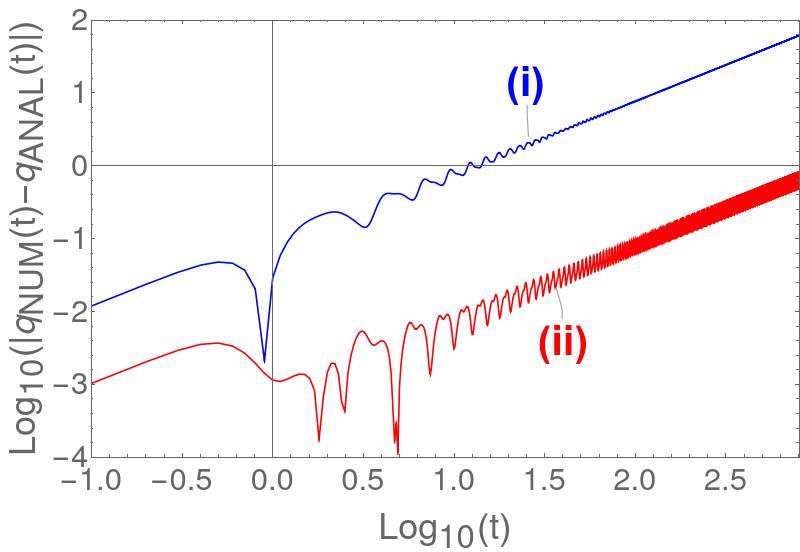}
\end{minipage}};
\end{tikzpicture}
}\:
}\\
\caption{Difference between the numerical solution $q(t)$ and the one produced by the method `analogous to Birkhoff' (in blue, (i)) or `analogous to Kolmogorov' (in red, (ii)).  From left to right for the cases 1) $\varepsilon=1\,$, $\omega_0=1$ and $\omega=1.002\,$, 2) $\varepsilon=1\,$, $\omega_0=1$ and $\omega=1.02\,$, 3) $\varepsilon=1\,$, $\omega_0=1$ and $\omega=1.2\,$.} 
\label{Numerical}
\end{figure}

Figure~\ref{Numerical} shows the difference (in $\mathrm{Log}_{10}\,$ scale) between the semi-analytical solutions obtained by methods (i) and (ii) as above and the numerical solution for the angle $q(t)\,$. Since the only systematically growing error is due to differences in the frequency estimates, all errors between the numerical and analytical solutions in the angle $q(t)$ grow linearly in time. We observe, however, that the Lindstedt method `analogous to Kolmogorov' always produces more precise results than the one `analogous to Birkhoff' with a difference in precision of about one order of magnitude when $d\omega$ is of order $10^{-3}$ and raising up to two orders of magnitude when $d\omega$ becomes of the order of unity.

\subsection{Hamiltonian preparation in the case of odd nonlinear couplings}

A particularity of the example treated above is the fact that the original Hamiltonian is analytic in the whole domain $J\in\mathbb{R}$. This changes, however, in more general models in which the Hamiltonian is of the form
$$
\Hscr(\vet{q},\vet{J}) = \vet{\omega_0}\cdot\vet{J}+ \varepsilon h(\vet{q},\vet{J};\varepsilon)
$$
where the development in powers of the variables $\vet{J}$ contains semi-integer powers, as is, for example, the case of polynomial nonlinear couplings containing odd terms in one or more of the oscillator variables $x_j,y_j$. Let us note that physical examples of such Hamiltonian systems, with oscillators non-linearly coupled through odd polynomial terms, are ubiquitous, and include the Fermi-Pasta-Ulam $\alpha-$model, the secular Hamiltonian in resonant cases of perturbed Keplerian $N-$body problems, magnetic bottle Hamiltonian models, etc. In the next section, we will present a formal algorithm applicable to a generic form for the function $\Hscr_1$. The sequence of normalizations in this algorithm is arranged so that the result agrees with the corresponding one obtained with the Lindstedt series. A particular example showing this agreement is presented in the Appendix.  

\section{KAM algorithm for isochronous systems}
\label{KAM}
We now give the Kolmogorov algorithm for generic isochronous systems with Hamiltonian
\begin{equation}
\label{h00}
\Hscr(\vet{q}, \vet{J})\,=\, \vet{\omega}_{0}\cdot\vet{J}+\sum_{i\geq 1}\sum_{s\geq \,3}\, \varepsilon^i \tilde{h}_{i,\, s}(\vet{q},\vet{J})
\end{equation}
where $\vet{q}\in\mathbb{T}^{n}\,$, $\vet{J}\in\mathcal{B}\subseteq\mathbb{R}^{n}\,$, $\tilde{h}_{i,\,s}=\mathcal{O}(||\vet{J}||^{\frac{s}{2}})\,$. 
We assume that the Hamiltonian~\eqref{h00} has a half-integer power dependence on $\vet{J}\,$, 
i.e. admits the expansion, for $n\rightarrow\infty$, of the truncated series
\begin{equation}
\begin{split}
\Hscr(\vet{q}, \vet{p}; \vet{J_0})\,&=\, \vet{\omega}_{0}\cdot\vet{J_0}+\vet{\omega}_{0}\cdot\vet{p}+\sum_{i\geq 1}\sum_{s\geq \,3}\, \varepsilon^i \tilde{h}_{i,\, s}(\vet{q},\vet{J_0}+\vet{p})\\
&=\, \vet{\omega}_{0}\cdot\vet{J_0}+\vet{\omega}_{0}\cdot\vet{p}+\sum_{i\geq 1}\sum_{s\geq \,3}\sum_{k= 0}^{n}\, \varepsilon^i \frac{\tilde{h}^{(k)}_{i,\, s}(\vet{q},\vet{J_0}+\vet{p})|_{\vet{p}=\vet{0}}}{k!}\,\vet{p}^{k}\, ,
\end{split}
\end{equation}
where $\tilde{h}^{(k)}$ are the $k-$th derivatives of $\tilde{h}$ with respect to $\vet{p}=\vet{J}-\vet{J_0}\,$. 

Apart from constants, the Hamiltonian can be written in the compact notation
\begin{equation}
\label{ham_0}
\Hscr(\vet{q},\vet{p}; \vet{J_{0}})\,=\, \vet{\omega}_{0}\cdot\vet{p}+\sum_{i\geq \,1}\,\varepsilon^i\, h_{i}(\vet{q},\vet{p}; \vet{J_{0}}) \,  .
\end{equation}

The algorithm allows to compute quasi-periodic orbits with a frequency $\vet{\omega}$ fixed in advance, given by
\begin{equation}
\label{omega}
\vet{\omega} = \vet{\omega}_{0} - \sum_{i\geq \,1} \, \varepsilon^i\, \vet{a}_{i}(\vet{J_{0}})\, , 
\end{equation}
where the parameters $\vet{J_{0}}$, whose values are to be specified in the end of the process, give the initial conditions for $\vet{J}$ of a trajectory on the torus with frequencies $\vet{\omega}\,$.  To this end, along with the normalizing canonical transformation, the algorith computes `on the go' the functions $\vet{a}_{i}(\vet{J}_{0})$ (`counter-terms'). 

Substituting the series~\eqref{omega} in the Hamiltonian \eqref{ham_0} we arrive at:
\begin{equation}
\label{ham_1}
\Hscr(\vet{q},\vet{p}; \vet{J_{0}})\,=\, \vet{\omega}\cdot\vet{p}+\sum_{i\geq \,1}\,\varepsilon^i\, h_{i}(\vet{q},\vet{p}; \vet{J_{0}}) + \sum_{i\geq \,1} \, \varepsilon^i\, \vet{a}_{i}(\vet{J_{0}})\cdot \vet{p}\, .
\end{equation}
In the following we use the notation
\begin{equation}
\label{forma_h}
\varepsilon^{i}\, h_{i}^{(k)}(\vet{q},\vet{p};\vet{J_{0}}) = \varepsilon^{i}\left(h_{i,0}^{(k)}(\vet{q};\vet{J_{0}})+h_{i,1}^{(k)}(\vet{q},\vet{p};\vet{J_{0}})+\sum_{j\geq\, 2} h_{i,j}^{(k)}(\vet{q},\vet{p};\vet{J_{0}})\right)\,,
\end{equation}
where $h_{i,0}^{(k)}$ is independent from $\vet{p}\,$, $h_{i,1}^{(k)}$ linear in $\vet{p}$ and the remaining sum is $\Oscr(\norm{\vet{p}}^2)\,$. To facilitate reading, the indices $(i,j,k)$ used in all subsequent expressions refer to
\begin{equation}
\label{notazione}
\begin{split}
&\phantom{h_{i,j}^{(k)} \,:}\,\, \: i=\textit{degree of the corresponding $\varepsilon$}\, , \\
&h_{i,j}^{(k)} \, : \: j= \textit{degree of $\vet{p}$} \, , \\
&\phantom{h_{i,j}^{(k)} \,:} \,  \: k= \textit{step of the algorithm} \, . 
\end{split}
\end{equation}
\noindent
We now have the following
\newtheorem{cnt.prop}{Proposition}
\setcounter{cnt.prop}{0}
\begin{cnt.prop}
Assume the vector $\vet{\omega}$ is non-resonant. There exist Lie generating functions $\chi_1^{(r)}$, $\chi_2^{(r)}$ such that, after $r$ normalization steps, the Hamiltonian~\eqref{ham_1} is given by the formal series:
\begin{align*}
 \Hscr ^{(r)}(\vet{q}^{(r)},\vet{p}^{(r)}; \vet{J}_{0}) 
  &= \bigg( \exp(L_{\varepsilon^r\chi_{2}^{(r)}})\big(\exp(L_{\varepsilon^r\chi_{1}^{(r)}})\Hscr ^{(r-1)}\big) \bigg)\bigg|_{\substack{\vet{q^{(r-1)}}=\vet{q}^{(r)} \\ \vet{p^{(r-1)}}=\vet{p}^{(r)}}}\, \\
  &=\,\vet{\omega}\cdot\vet{p}+\varepsilon\,C_{1}+\ldots + \varepsilon^{r-1}\,C_{r-1}+\varepsilon^{r}\,C_{r}\\
 &\phantom{=}+\varepsilon\,h_{1}^{(r)}(\vet{q},\vet{p}; \vet{J_{0}}) +\ldots + \varepsilon^{r} \,h_{r}^{(r)}(\vet{q},\vet{p}; \vet{J_{0}})  \\
&\phantom{=}+ \varepsilon^{r+1} \,h_{r+1}^{(r)}(\vet{q},\vet{p}; \vet{J_{0}})+ \varepsilon^{r+1}\,\vet{a}_{r+1}(\vet{J_{0}})\cdot\vet{p}+\ldots  \, ,
 \end{align*} 
 where $h_{j}^{(r)}=\sum_{i\geq 0} h_{j,i}^{(r)}\,\forall\,j\geq r+1\,$ and $h_{j}^{(r)}=\sum_{i\geq 2} h_{j,i}^{(r)}\,\in\mathcal{O}(||\vet{p}||^{2})\,\forall\,j=1,\ldots,r$, with
\begin{align*}
&h_{k,0}^{(r)}=C_{k}=\avg{h_{k,0}^{(k-1)}}-\vet{\omega}\cdot\vet{K}^{(k)}\,& &\;\;\forall \; 1\leq k\leq r\, ,\\
&h_{k,0}^{(r)}=h_{k,0}^{(r-1)}
& &\;\;
\forall\;  r+1\leq k\leq 2r-1\, ,\\[1.4ex]
&h_{2r,0}^{(r)}= \sum_{j=0}^{\lfloor \frac{2r-1}{r} \rfloor} \frac{1}{j!}L_{\chi_{1}^{(r)}(\vet{q})}^{j}\left(h_{2r-jr,j}^{(r-1)} \right)  & &\\
&h_{k,0}^{(r)}=\sum_{j=0}^{\lfloor \frac{k-1}{r} \rfloor -2}\sum_{s=0}^{\lfloor \frac{k-jr-1}{r} \rfloor } \frac{1}{j!\,s!}L_{\chi_{2}^{(r)}(\vet{q},\vet{p})}^{j}L_{\chi_{1}^{(r)}(\vet{q})}^{s}\left(h_{k-jr-sr,s}^{(r-1)} \right) &  & \notag \\
&\qquad +\frac{1}{\left(\lfloor \frac{k-1}{r} \rfloor -1 \right)!}L_{\chi_{2}^{(r)}(\vet{q},\vet{p})}^{\lfloor (k-1)/r \rfloor -1}\left(h_{k-\left(\lfloor \frac{k-1}{r} \rfloor -1\right) r ,0}^{(r-1)} \right)  &  &\;\;\forall \;  k\geq 2r+1\,  , k\neq mr\,, m\in\mathbb{N}\, ,\notag\\[1.6ex]
& h_{k,0}^{(r)} = \sum_{j=0}^{\lfloor \frac{k-1}{r} \rfloor-1 }\sum_{s=0}^{\lfloor \frac{k-jr-1}{r} \rfloor } \frac{1}{j!\,s!}L_{\chi_{2}^{(r)}(\vet{q},\vet{p})}^{j}L_{\chi_{1}^{(r)}(\vet{q})}^{s}\left(h_{k-jr-sr,s}^{(r-1)} \right) 
 &  &\;\;\forall\; k\geq 2r+1\, ,k= mr\,, m\in\mathbb{N}\, ,\\[1.4ex]
& h_{k,1}^{(r)}=0 &
&\;\;\forall\; 1\leq k \leq r\, ,\\[1.4ex]
& h_{k,1}^{(r)} = \sum_{j=0}^{\lfloor \frac{k-1}{r} \rfloor-1 }\sum_{s=0}^{\lfloor \frac{k-jr-1}{r} \rfloor } \frac{1}{j!\,s!}L_{\chi_{2}^{(r)}(\vet{q},\vet{p})}^{j}L_{\chi_{1}^{(r)}(\vet{q})}^{s}\left(h_{k-jr-sr,1+s}^{(r-1)} \right) 
 &  &\;\;\forall \;  k\geq r+1\, , k\neq m\,r\, ,\; m\in\mathbb{N}\, , \\[1.4ex] 
& h_{k,1}^{(r)}=\sum_{j=0}^{\lfloor \frac{k-2}{r} \rfloor }\sum_{s=0}^{\lfloor \frac{k-jr-1}{r} \rfloor } \frac{1}{j!\,s!}L_{\chi_{2}^{(r)}(\vet{q},\vet{p})}^{j}L_{\chi_{1}^{(r)}(\vet{q})}^{s}\left(h_{k-jr-sr,1+s}^{(r-1)} \right) & & \\
&\quad\phantom{h_{k,i}^ {(r)}=} + \frac{m-1}{m!} L_{\chi_{2}^{(r)}(\vet{q},\vet{p})}^{m-1}\left(h_{r,1}^{(r-1)}\right) & & \; \;\forall \;  k\geq r+1\, , k= m\,r\, , \;  m\in\mathbb{N}\, , \\[1.4ex]
&h_{k,i}^{(r)}=h_{k,i}^{(r-1)}
& &\;\;\forall \; 1\leq k\leq r\, ,\; i\geq 2\, ,\\[1,4ex]
& h_{k,i}^{(r)} = \sum_{j=0}^{\lfloor \frac{k-1}{r} \rfloor }\sum_{s=0}^{\lfloor \frac{k-jr-1}{r} \rfloor } \frac{1}{j!\,s!}L_{\chi_{2}^{(r)}(\vet{q},\vet{p})}^{j}L_{\chi_{1}^{(r)}(\vet{q})}^{s}\left(h_{k-jr-sr,i+s}^{(r-1)} \right) 
 &  &\;\;\forall \;  k\geq r+1\, , \; i\geq 2\, .
\end{align*}
Given the Fourier series 
$\displaystyle{h_{r,0}^{(r-1)}(\vet{q})=\sum_{\vet{k}\in\mathbb{Z}^n} c_{\vet{k}}^{(r-1)}e^{i \vet{k}\cdot\vet{q}}}$ and $\displaystyle{h_{r,1}^{(r-1)}(\vet{q},\vet{p})=\sum_{|\vet{j}|=1}\sum_{\vet{k}\in\mathbb{Z}^n}c_{\vet{j}\vet{k}}^{(r)}\vet{p}^{\vet{j}}e^{i \vet{k}\cdot\vet{q}}}\,$,
the generating functions $\chi_{1}^{(r)}$, $\chi_{2}^{(r)}$ are defined by
\begin{align*}
&\chi_{1}^{(r)}(\vet{q})=X^{(r)}(\vet{q})+\vet{K}^{(r)}\cdot\vet{q}\,=\sum_{\vet{k}\in\mathbb{Z}^n\setminus\lbrace\vet{0}\rbrace} \left(\frac{c^{(r-1)}_{\vet{k}}}{i\vet{k}\cdot\vet{\omega}} \, e^{i \vet{k}\cdot\vet{q}}-\frac{c^{(r-1)}_{\vet{k}}}{\vet{k}\cdot\vet{\omega}}\,\vet{k}\cdot\vet{q}\right)\, , \\
&\chi_{2}^{(r)}(\vet{q},\vet{p})=\tilde{\chi}_{2}^{(r)}(\vet{q},\vet{p})+ \vet{S}^{(r)}\cdot\vet{p}\\
&\phantom{\chi_{2}^{(r)}(\vet{q},\vet{p})}=\vet{Y}^{(r)}(\vet{q})\cdot\vet{p}+ \vet{S}^{(r)}\cdot\vet{p}=\sum_{|\vet{j}|=1}\sum_{\vet{k}\in\mathbb{Z}^n\setminus\lbrace\vet{0}\rbrace}\left( \frac{c_{\vet{j}\vet{k}}^{(r)}}{i\vet{k}\cdot\vet{\omega}} \, e^{i \vet{k}\cdot\vet{q}}+\frac{i\,c_{\vet{j}\vet{k}}^{(r)}}{\vet{k}\cdot\vet{\omega}}\right)\vet{p}^{\vet{j}}\,.
\end{align*}
\end{cnt.prop}
\noindent
{\bf Proof of the proposition.}
The generic $r$-th iterative step of the algorithm is defined as follows: after $r-1$ steps, the hamiltonian~\eqref{ham_1} has the form:
 \begin{equation}
\label{ham_r-1}
\begin{split}
\Hscr^{(r-1)}(\vet{q},\vet{p}; \vet{J_{0}})\,=\, &\vet{\omega}\cdot\vet{p}+\varepsilon\,C_{1}+\ldots + \varepsilon^{r-1}\,C_{r-1}\\
&+\varepsilon\,h_{1}^{(r-1)}(\vet{q},\vet{p}; \vet{J_{0}}) +\ldots + \varepsilon^{r-1} \,h_{r-1}^{(r-1)}(\vet{q},\vet{p}; \vet{J_{0}})  \\
&+ \varepsilon^{r} \,h_{r}^{(r-1)}(\vet{q},\vet{p}; \vet{J_{0}}) + \varepsilon^{r}\,\vet{a}_{r}(\vet{J_{0}})\cdot \vet{p} \\
&+ \varepsilon^{r+1} \,h_{r+1}^{(r-1)}(\vet{q},\vet{p}; \vet{J_{0}})    + \varepsilon^{r+1}\,\vet{a}_{r+1}(\vet{J_{0}})\cdot\vet{p} + \ldots\, ,
\end{split}
\end{equation}
where $C_{1}\,\ldots\, C_{r-1}$ are constants and $h_{i}^{(r-1)}\in \Oscr(\norm{\vet{p}}^{2})\,$ $\forall\,i\,=\,1,\ldots,r-1\,$. 
Taking into account the notation~\eqref{notazione}, we re-define the quantity $h_{i}^{(k)}$ according to 
\begin{equation}
\label{new.def}
 h_{i,1}^{(k)} \leftarrow h_{i,1}^{(k)} + \vet{a}_{i}\cdot\vet{p} = \left(\nabla_{\vet{p}} h_{i,1}^{(k)} +\vet{a}_{i} \right) \cdot\vet{p}\quad \forall \, i\geq r \, .
 \end{equation} 
We then have:

\emph{\textbf{First part of the proof}:} the $r$-th normalization step consists of two substeps, each involving a canonical transformation. 
\begin{itemize}
\item 
\textit{First half step:} we set 
\begin{equation}
\label{serie.lie.h.hhat}
\widehat{\Hscr}^{(r)} = \exp(L_{\varepsilon^{r}\,\chi_{1}^{(r)}(\vet{q})})\Hscr ^{(r-1)}= \sum_{j\geq 0} \frac{(\varepsilon^{r})^{j}}{j!} L_{\chi_{1}^{(r)}(\vet{q})}^{j}\Hscr ^{(r-1)}\,,
\end{equation}
\end{itemize}
where the generating function $\chi_{1}^{(r)}$ is defined as $\chi_{1}^{(r)}(\vet{q})=X^{(r)}(\vet{q})+\vet{K}^{(r)}\cdot\vet{q}\,$, with $\vet{K}^{(r)}$ an appropriate constant vector defined below, and 
$X^{(r)}$ is defined through the homological equation:
\begin{equation}
\label{uccidi.term.q}
\bigpoisson {\vet{\omega}\cdot \vet{p}}{X^{(r)}(\vet{q})} + h_{r, 0}^{(r-1)}(\vet{q}) = \avg{h_{r, 0}^{(r-1)}(\vet{q})}\, ,
\end{equation}
where $\avg{.}$ denotes the mean over $\vet{q}\,$.
The function $X^{(r)}(\vet{q})$ eliminates all terms of order $\Oscr(\varepsilon^r)\,$ depending only on the angles $\vet{q}\,$ in the Hamiltonian  $\Hscr^{(r-1)}$. Writing $h_{r,0}^{(r-1)}(\vet{q})$ in the Fourier form 
$ \displaystyle{h_{r,0}^{(r-1)}(\vet{q})=\sum_{\vet{k}\in\interi^n} c_{\vet{k}}^{(r-1)}e^{i \vet{k}\cdot\vet{q}}}$ from~\eqref{uccidi.term.q} we find:
\begin{equation}
\label{def.X}
 X^{(r)}(\vet{q})=\sum_{\vet{k}\in\interi^n\setminus\lbrace\vet{0}\rbrace} \frac{c^{(r-1)}_{\vet{k}}}{i\vet{k}\cdot\vet{\omega}} \, e^{i \vet{k}\cdot\vet{q}}. \, 
\end{equation}
We then impose the condition that the terms of order $\Oscr(\varepsilon^{r})$ linear in $\vet{p}$ have zero \begin{equation}
 \label{media.nulla}
 \avg{h_{r,1}^{(r-1)}(\vet{q},\vet{p};\vet{J_{0}})} =0\, .
\end{equation}
This specifies $\vet{a}_{r}$ as a function of $\vet{J}_{0}\,$ via Eq~\eqref{new.def}.

Finally, we specify the constant vector $\vet{K}^{(r)}\,$. To this end, we impose the condition that the solution in $\vet{p}$ has the form $\sum_{\vet{k}} A_{\vet{k}}\left(\cos(\vet{k}\cdot\vet{q})-1\right)\,$, so that, at time $t=0$, we have $p(0)=0$. Writing~\eqref{def.X} as
 \begin{equation}
 \label{riscrittura_X}
 X^{(r)}(\vet{q})=\sum_{\vet{k}\in\interi^n\setminus\lbrace\vet{0}\rbrace}\left[- \frac{i\,c^{(r-1)}_{\vet{k}}}{\vet{k}\cdot\vet{\omega}}\cos(\vet{k}\cdot\vet{q}) + \frac{c^{(r-1)}_{\vet{k}}}{\vet{k}\cdot\vet{\omega}}\sin(\vet{k}\cdot\vet{q})  \right]\, ,
\end{equation}
the poisson bracket of $\chi_{1}^{(r)}(\vet{q})$ with the i-th component $p_i\,$ ($1 \leq \, i \, \leq n\,$) of the vector $\vet{p}$ yields the following expressions for the i-th components $K_i^{(r)}$ and $k_i$ of the vectors $\vet{K}^{(r)}$ and $\vet{k}\,$:
\begin{equation}
\begin{split}
\poisson{p_i\,}{X^{(r)}(\vet{q})}+\poisson{p_i\,}{\vet{K}^{(r)}\cdot\vet{q}}&= \sum_{\vet{k}\in\interi^n\setminus\lbrace\vet{0}\rbrace}\left[- \bigpoisson{p_i}{\frac{i\,c^{(r-1)}_{\vet{k}}}{\vet{k}\cdot\vet{\omega}}\cos(\vet{k}\cdot\vet{q})} +\bigpoisson{p_i}{\frac{c^{(r-1)}_{\vet{k}}}{\vet{k}\cdot\vet{\omega}}\sin(\vet{k}\cdot\vet{q})} \right]-K^{(r)}_i \\
&=-\sum_{\vet{k}\in\interi^n\setminus\lbrace\vet{0}\rbrace}\left[\,\frac{i\,c^{(r-1)}_{\vet{k}}}{\vet{k}\cdot\vet{\omega}}\,k_i\,\sin(\vet{k}\cdot\vet{q}) + \frac{c^{(r-1)}_{\vet{k}}}{\vet{k}\cdot\vet{\omega}}\, k_i\,\cos(\vet{k}\cdot\vet{q}) \, \right]-K^{(r)}_i=\\
&=-\sum_{\vet{k}\in\interi^n\setminus\lbrace\vet{0}\rbrace}\frac{i\,c^{(r-1)}_{\vet{k}}}{\vet{k}\cdot\vet{\omega}}\,k_i\,\sin(\vet{k}\cdot\vet{q}) - \sum_{\vet{k}\in\interi^n\setminus\lbrace\vet{0}\rbrace}\frac{c^{(r-1)}_{\vet{k}}}{\vet{k}\cdot\vet{\omega}}\, k_i\left(\cos(\vet{k}\cdot\vet{q})-1\right) \, .
\end{split}
\end{equation}  
Hence
\begin{equation}
K^{(r)}_i=-\sum_{\vet{k}\in\interi^n\setminus\lbrace\vet{0}\rbrace}\frac{c^{(r-1)}_{\vet{k}}}{\vet{k}\cdot\vet{\omega}}\, k_i\, ,
\end{equation} 
i.e.,
\begin{equation}
\label{def_cost}
\vet{K}^{(r)}=-\sum_{\vet{k}\in\interi^n\setminus\lbrace\vet{0}\rbrace}\frac{c^{(r-1)}_{\vet{k}}}{\vet{k}\cdot\vet{\omega}}\, \vet{k}\, .
\end{equation}
\\Finally, we compute $\widehat{\Hscr}^{(r)}$ in~\eqref{serie.lie.h.hhat} as:
 \begin{equation}
\label{ham_hat_r}
\begin{split}
\widehat{\Hscr}^{(r)}(\vet{q},\vet{p}; \vet{J_{0}})\,=\, &\vet{\omega}\cdot\vet{p}+\varepsilon\,C_{1}+\ldots + \varepsilon^{r-1}\,C_{r-1}+\varepsilon^{r}\,C_{r}\\
&+\varepsilon\,\widehat{h}_{1}^{(r)}(\vet{q},\vet{p}; \vet{J_{0}}) +\ldots + \varepsilon^{r-1} \,\widehat{h}_{r-1}^{(r)}(\vet{q},\vet{p}; \vet{J_{0}})  \\
&+ \varepsilon^{r} \,\widehat{h}_{r}^{(r)}(\vet{q},\vet{p}; \vet{J_{0}}) + \varepsilon^{r+1} \,\widehat{h}_{r+1}^{(r)}(\vet{q},\vet{p}; \vet{J_{0}})  +\ldots  \, , 
\end{split}
\end{equation}
where $\widehat{h}_{j}^{(r)}=\sum_{i\geq 0} \widehat{h}_{j,i}^{(r)}\,\forall\,j\geq r\,$ and $\widehat{h}_{j}^{(r)}=\sum_{i\geq 2} \widehat{h}_{j,i}^{(r)}\in\Oscr(\norm{\vet{p}}^{2})\,$ $\forall\,j=1,\ldots,r-1\,$, where:
\begin{align}
\label{0_12}
&\widehat{h}_{k,0}^{(r)}=C_{k}=\avg{h_{k,0}^{(k-1)}}-\vet{\omega}\cdot\vet{K}^{(k)}& &\forall\,1\leq k\leq r\, ,\\
\label{0_3}
&\widehat{h}_{k,0}^{(r)}=h_{k,0}^{(r-1)}& 
&\forall\: r+1\leq k\leq 2r-1\, ,  \\
\label{0_4}
& \widehat{h}_{k,0}^{(r)}=\sum_{j=0}^{\lfloor \frac{k-1}{r} \rfloor} \frac{1}{j!}L_{\chi_{1}^{(r)}(\vet{q})}^{j}\left(h_{k-jr,j}^{(r-1)} \right) & &\forall\, k\geq 2r\, \\
\label{1_0}
&\widehat{h}_{k,1}^{(r)}=0 & &\forall \: 1\leq k\leq r-1\, ,\\
\label{1_1}
&\widehat{h}_{r,1}^{(r)}=h_{r,1}^{(r-1)}\, ,& 
& \\
\label{1_2}
& \widehat{h}_{k,i}^{(r)}=\sum_{j=0}^{\lfloor \frac{k-1}{r} \rfloor} \frac{1}{j!}L_{\chi_{1}^{(r)}(\vet{q})}^{j}\left(h_{k-jr,i+j}^{(r-1)} \right) 
& &\forall \:  k\geq r+1\, , \: i\geq 1\, ,\\
\label{2_1}
&\widehat{h}_{k,i}^{(r)}=h_{k,i}^{(r-1)}&
&\forall \: 1\leq k\leq r\, , \:  i\geq 2\, .
\end{align}
\begin{itemize}
\item 
\textit{Second half step (of r-th):} We compute
\begin{equation}
\label{serie.lie.h.r}
\Hscr^{(r)} = \exp(L_{\varepsilon^{r}\,\chi_{2}^{(r)}(\vet{q},\vet{p})})\widehat{\Hscr} ^{(r)}= \sum_{j\geq 0} \frac{(\varepsilon^{r})^{j}}{j!} L_{\chi_{2}^{(r)}(\vet{q},\vet{p})}^{j}\widehat{\Hscr} ^{(r)}\,,
\end{equation}
\end{itemize}
with a generating function $\chi_{2}^{(r)}(\vet{q},\vet{p})$ linear in $\vet{p}$. Setting $\chi_{2}^{(r)}(\vet{q},\vet{p})=\tilde{\chi}_{2}^{(r)}(\vet{q},\vet{p})+ \vet{S}^{(r)}\cdot\vet{p}=\vet{Y}^{(r)}(\vet{q})\cdot\vet{p}+ \vet{S}^{(r)}\cdot\vet{p}\,$. The function $\tilde{\chi}_{2}^{(r)}(\vet{q},\vet{p})$ satisfies the homological equation 
\begin{equation}
\label{uccidi.term.p}
\bigpoisson {\vet{\omega}\cdot \vet{p}}{\tilde{\chi}_{2}^{(r)}(\vet{q},\vet{p})} + \widehat{h}_{r, 1}^{(r)}(\vet{q},\vet{p}) = 0\,.
\end{equation}
Setting $\displaystyle{\widehat{h}_{r,1}^{(r)}(\vet{q},\vet{p})=\sum_{|\vet{j}|=1}\sum_{\vet{k}\in\interi^n}c_{\vet{j}\vet{k}}^{(r)}\vet{p}^{\vet{j}}e^{i \vet{k}\cdot\vet{q}}}$ the solution of (\ref{uccidi.term.p}) is
\begin{equation}\label{def.chi}
 \displaystyle{ \tilde{\chi}_{2}^{(r)}(\vet{q},\vet{p})=\sum_{|\vet{j}|=1}\sum_{\vet{k}\in\interi^n\setminus\lbrace\vet{0}\rbrace} \frac{c_{\vet{j}\vet{k}}^{(r)}}{i\vet{k}\cdot\vet{\omega}} \vet{p}^{\vet{j}}\, e^{i \vet{k}\cdot\vet{q}}}. \, 
\end{equation}

We finally compute the constant vector $\vet{S}^{(r)}\,$ by the condition $\vet{q}(0)=0$ at the time $t=0$. By the poisson bracket
 \begin{equation}
 \begin{split}
\poisson{q_i\,}{\chi_{2}^{(r)}}&=\poisson{q_i\,}{\tilde{\chi}_{2}^{(r)}(\vet{q}, \vet{p})}+\poisson{q_i\,}{\vet{S}^{(r)}\cdot\vet{p}}\\
&= \sum_{|\vet{j}|=1}\left(\sum_{\vet{k}\in\interi^n\setminus\lbrace\vet{0}\rbrace}\left[ \bigpoisson{q_i}{\frac{c_{\vet{j}\vet{k}}^{(r)}}{i\,\vet{k}\cdot\vet{\omega}}\,\vet{p}^{\vet{j}}\,\cos(\vet{k}\cdot\vet{q})} +\bigpoisson{q_i}{\frac{c^{(r)}_{\vet{j}\vet{k}}}{\vet{k}\cdot\vet{\omega}}\,\vet{p}^{\vet{j}}\sin(\vet{k}\cdot\vet{q})} \right]+\bigpoisson{q_i\,}{S_{\vet{j}}^{(r)}\vet{p}^{\vet{j}}}\right)\\
&=\sum_{|\vet{j}|=1}\left(\sum_{\vet{k}\in\interi^n\setminus\lbrace\vet{0}\rbrace}\left[\frac{c_{\vet{j}\vet{k}}^{(r)}}{i\,\vet{k}\cdot\vet{\omega}}\,j_i\,p_i^{j_i-1}\,\cos(\vet{k}\cdot\vet{q}) +\frac{c^{(r)}_{\vet{j}\vet{k}}}{\vet{k}\cdot\vet{\omega}}\,j_i\,p_i^{j_i-1}\sin(\vet{k}\cdot\vet{q}) \right]+{S_{\vet{j}}^{(r)}\,j_i\,p_i^{j_i-1}}\right)\\
&=\sum_{|\vet{j}|=1}\sum_{\vet{k}\in\interi^n\setminus\lbrace\vet{0}\rbrace}\left[ \frac{c_{\vet{j}\vet{k}}^{(r)}}{i\,\vet{k}\cdot\vet{\omega}}\,j_i\,p_i^{j_i-1}\,\left(\cos(\vet{k}\cdot\vet{q})-1\right) +\frac{c^{(r)}_{\vet{j}\vet{k}}}{\vet{k}\cdot\vet{\omega}}\,j_i\,p_i^{j_i-1}\sin(\vet{k}\cdot\vet{q}) \right] \, 
\end{split}
\end{equation} 
we obtain
\begin{equation}
\label{def_costante_2}
S^{(r)}_\vet{j}=\sum_{\vet{k}\in\interi^n\setminus\lbrace\vet{0}\rbrace}\frac{i\,c^{(r)}_{\vet{j}\vet{k}}}{\vet{k}\cdot\vet{\omega}}\, .
\end{equation} 

The Hamiltonian $\Hscr^{(r)}$ (Eq.~\eqref{serie.lie.h.r}) is:
 \begin{equation}
\label{ham_hat_rb}
\begin{split}
\Hscr^{(r)}(\vet{q},\vet{p}; \vet{J_{0}})\,=\, &\vet{\omega}\cdot\vet{p}+\varepsilon\,C_{1}+\ldots + \varepsilon^{r-1}\,C_{r-1}+\varepsilon^{r}\,C_{r}\\
&+\varepsilon\,h_{1}^{(r)}(\vet{q},\vet{p}; \vet{J_{0}}) +\ldots + \varepsilon^{r} \,h_{r}^{(r)}(\vet{q},\vet{p}; \vet{J_{0}})  \\
&+ \varepsilon^{r+1} \,h_{r+1}^{(r)}(\vet{q},\vet{p}; \vet{J_{0}})+\ldots  \, , 
\end{split}
\end{equation}
where $h_{j}^{(r)}=\sum_{i\geq 0} h_{j,i}^{(r)}\,\forall\,j\geq r+1\,$ and $h_{j}^{(r)}=\sum_{i\geq 2} h_{j,i}^{(r)}\,\in\Oscr(\norm{\vet{p}}^{2})\,\forall\,j=1,\ldots,r\,$, where:
\begin{align}
\label{hat0_1}
&h_{k,0}^{(r)}=\widehat{h}_{k,0}^{(r)} &
&\;\;\forall\; 1\leq k\leq 2r\, , \\
\label{hat0_2}
& h_{k,0}^{(r)}=\sum_{j=0}^{\lfloor \frac{k-1}{r} \rfloor} \frac{1}{j!}L_{\chi_{2}^{(r)}(\vet{q},\vet{p})}^{j}\left(\widehat{h}_{k-jr,0}^{(r)} \right) &
&\;\;\forall\; k\geq 2r+1\, ,\\
\label{hat1_0}
& h_{k,1}^{(r)}=0 &
&\;\;\forall\; 1\leq k \leq r\, ,\\
\label{hat1_1NO}
& h_{k,1}^{(r)}=\sum_{j=0}^{\lfloor \frac{k-1}{r} \rfloor} \frac{1}{j!}L_{\chi_{2}^{(r)}(\vet{q},\vet{p})}^{j}\left(\widehat{h}_{k-jr,1}^{(r)} \right) &
&\;\;\forall \;  k\geq r+1\, , k\neq m\,r\,  (m\in\mathbb{N})\, ,\\
\label{hat1_1SI}
& h_{k,1}^{(r)}=\sum_{j=0}^{\lfloor \frac{k-1}{r} \rfloor} \frac{1}{j!}L_{\chi_{2}^{(r)}(\vet{q},\vet{p})}^{j}\left(\widehat{h}_{k-jr,1}^{(r)} \right)+ \frac{1}{m!} L_{\chi_{2}^{(r)}(\vet{q},\vet{p})}^{m}\left(\vet{\omega}\cdot\vet{p}\right) &
& \;\;\forall \;  k\geq r+1\, , k= m\,r\, (m\in\mathbb{N})\, .\\
\label{hat2_1}
&h_{k,i}^{(r)}=\widehat{h}_{k,i}^{(r)} & 
&\;\;\forall \; 1\leq k\leq r\, , \:  i\geq 2\, ,\\
\label{hat2_2}
& h_{k,i}^{(r)}=\sum_{j=0}^{\lfloor \frac{k-1}{r} \rfloor} \frac{1}{j!}L_{\chi_{2}^{(r)}(\vet{q},\vet{p})}^{j}\left(\widehat{h}_{k-jr,i}^{(r)} \right) &
&\;\;\forall \;  k\geq r+1\, , \: i\geq 2\, .
\end{align}
Equation~\eqref{hat1_1SI}, using homological equation~\eqref{uccidi.term.p}, can be written equivalently as:
\begin{align}
\label{hat1_1SI_bis}
& h_{k,1}^{(r)}=\sum_{j=0}^{m-2} \frac{1}{j!}L_{\chi_{2}^{(r)}(\vet{q},\vet{p})}^{j}\left(\widehat{h}_{k-jr,1}^{(r)} \right)  + \frac{m-1}{m!} L_{\chi_{2}^{(r)}(\vet{q},\vet{p})}^{m-1}\left(\widehat{h}_{r,1}^{(r-1)}\right) & &\;\;\forall \;  k\geq r+1\, , k= m\,r\, (m\in\mathbb{N})\, 
\end{align}
where, from~\eqref{hat1_1SI} we have
\begin{align*}
&\sum_{j=0}^{m-1} \frac{1}{j!}L_{\chi_{2}^{(r)}(\vet{q},\vet{p})}^{j}\left(\widehat{h}_{mr-jr,1}^{(r)} \right) + \frac{1}{m!} L_{\chi_{2}^{(r)}(\vet{q},\vet{p})}^{m}\left(\vet{\omega}\cdot\vet{p}\right)=\\
&\sum_{j=0}^{m-2} \frac{1}{j!}L_{\chi_{2}^{(r)}(\vet{q},\vet{p})}^{j}\left(\widehat{h}_{mr-jr,1}^{(r)} \right)+\frac{1}{(m-1)!}L_{\chi_{2}^{(r)}(\vet{q},\vet{p})}^{m-1}\left(\widehat{h}_{r,1}^{(r)} \right) + \frac{1}{m!} L_{\chi_{2}^{(r)}(\vet{q},\vet{p})}^{m}\left(\vet{\omega}\cdot\vet{p}\right)=\\
&\sum_{j=0}^{m-2} \frac{1}{j!}L_{\chi_{2}^{(r)}(\vet{q},\vet{p})}^{j}\left(\widehat{h}_{mr-jr,1}^{(r)} \right)+\frac{1}{m!}L_{\chi_{2}^{(r)}(\vet{q},\vet{p})}^{m-1}\left(\widehat{h}_{r,1}^{(r)}  +  L_{\chi_{2}^{(r)}(\vet{q},\vet{p})}\left(\vet{\omega}\cdot\vet{p}\right)\right)+\frac{m-1}{m!}\,L_{\chi_{2}^{(r)}(\vet{q},\vet{p})}^{m-1}\left(\widehat{h}_{r,1}^{(r)} \right).
\end{align*}
 
\emph{\textbf{Second part of the proof}:} using the formulas~\eqref{0_12}--\eqref{2_1} and~\eqref{hat0_1}--\eqref{hat2_2}, we can express each term in the function $h^{(r)}$ in terms of the functions $h^{(r-1)}\,$ instead of $\widehat{h}^{(r)}\,$. From Eq~\eqref{hat0_1} we have
\begin{align}
&h_{k,0}^{(r)}=\widehat{h}_{k,0}^{(r)}=C_{k}=\avg{h_{k,0}^{(k-1)}}-\vet{\omega}\cdot\vet{K}^{(k)}\, \qquad~\forall \; 1\leq k\leq r\, ,\\
&h_{k,0}^{(r)}=\widehat{h}_{k,0}^{(r)}=h_{k,0}^{(r-1)} 
\qquad~~~~~~~~~~~~~~~~~~~~~~~~~~~~~~\forall \: r+1\leq k\leq 2r-1\, ,\\
&h_{2r,0}^{(r)}=\widehat{h}_{2r,0}^{(r)}=  \sum_{j=0}^{\lfloor \frac{2r-1}{r} \rfloor} \frac{1}{j!}L_{\chi_{1}^{(r)}(\vet{q})}^{j}\left(h_{2r-jr,j}^{(r-1)} \right)=h_{2r,0}^{(r-1)}+ L_{\chi_{1}^{(r)}(\vet{q})}\left(h_{r,1}^{(r-1)}\right)\,,
\end{align}
where we used Eqs~\eqref{0_12},~\eqref{0_3} and~\eqref{0_4}, respectively.
Using~\eqref{2_1} we have
\begin{equation}
h_{k,i}^{(r)}=\widehat{h}_{k,i}^{(r)}=h_{k,i}^{(r-1)}\quad\forall \: 1\leq k\leq r\, , \:  i\geq 2\, .
\end{equation}
Equation~\eqref{hat2_2} can now be written in the form
\begin{equation}
\label{primo_split}
h_{k,i}^{(r)}=\sum_{j=0}^{\lfloor \frac{k-1}{r} \rfloor -1} \frac{1}{j!}L_{\chi_{2}^{(r)}(\vet{q},\vet{p})}^{j}\left(\widehat{h}_{k-jr,i}^{(r)} \right) + \frac{1}{\lfloor \frac{k-1}{r} \rfloor!}L_{\chi_{2}^{(r)}(\vet{q},\vet{p})}^{\lfloor (k-1)/r \rfloor}\left(\widehat{h}_{k-\lfloor \frac{k-1}{r} \rfloor r,i}^{(r)} \right)
\end{equation}
$ \forall \:  k\geq r+1\, , \: i\geq 2\, $. The indices $i,j,k$ in the first term of the above equation satisfy the relation
\begin{equation}
\label{1gruppo}
k-jr\geq k-\lfloor \frac{k-1}{r} \rfloor r + r \geq r+1\, ,
\end{equation}
where we have used the inequalities $j\leq \lfloor (k-1)/r \rfloor -1 $ and $\lfloor k-1 \rfloor \leq k-1\,$. Then Eq~\eqref{1gruppo} ensures that the first term of~\eqref{primo_split} satisfies the definition~\eqref{1_2}. For the second term in (\ref{primo_split}), we have the following useful (also in the sequel) remark: we can write $k=mr+f\,$, where $0\leq f\leq r-1\,$ and $m\in\mathbb{N}\,$. Thus
\begin{align}
k-\lfloor \frac{k-1}{r} \rfloor r = mr + f - \lfloor \frac{mr + f-1}{r} \rfloor r =f -\lfloor \frac{f-1}{r} \rfloor r\, .
\end{align}
Then
\begin{align}
\label{2gruppo_multiplo}
k-\lfloor \frac{k-1}{r} \rfloor r =f -\lfloor \frac{f-1}{r} \rfloor r= r\quad \mathrm{if}\; \, f=0\, 
\end{align}
and
\begin{align}
\label{2gruppo_no_multiplo}
1\leq k-\lfloor \frac{k-1}{r} \rfloor r =f -\lfloor \frac{f-1}{r} \rfloor r \leq r-1\quad\mathrm{if}\;\, 1\leq f \leq r-1\, .
\end{align}
We can conclude that $1\leq k-\lfloor (k-1)/r \rfloor r \leq r\,$, implying that the definition~\eqref{2_1} holds for the second term in (\ref{primo_split}). We can then write~\eqref{primo_split} as\footnote{Observe that, due to~\eqref{2gruppo_multiplo} and~\eqref{2gruppo_no_multiplo}, we have that if $j= \lfloor (k-1)/r \rfloor\,$, then $0\leq \lfloor(k-jr-1)/r \rfloor=\lfloor(k-\lfloor(k-1)/r \rfloor r -1)/r \rfloor \leq\lfloor (r-1)/r \rfloor = 0\, $, i.e., $s=0\,$. This allows to join all terms in a single sum.}
\begin{align}
h_{k,i}^{(r)}&=\sum_{j=0}^{\lfloor \frac{k-1}{r} \rfloor -1}\sum_{s=0}^{\lfloor \frac{k-jr-1}{r} \rfloor } \frac{1}{j!\,s!}L_{\chi_{2}^{(r)}(\vet{q},\vet{p})}^{j}L_{\chi_{1}^{(r)}(\vet{q})}^{s}\left(h_{k-jr-sr,i+s}^{(r-1)} \right) + \frac{1}{\lfloor \frac{k-1}{r} \rfloor!}L_{\chi_{2}^{(r)}(\vet{q},\vet{p})}^{\lfloor (k-1)/r \rfloor}\left(h_{k-\lfloor \frac{k-1}{r} \rfloor r,i}^{(r-1)} \right)\\
&=\sum_{j=0}^{\lfloor \frac{k-1}{r} \rfloor }\sum_{s=0}^{\lfloor \frac{k-jr-1}{r} \rfloor } \frac{1}{j!\,s!}L_{\chi_{2}^{(r)}(\vet{q},\vet{p})}^{j}L_{\chi_{1}^{(r)}(\vet{q})}^{s}\left(h_{k-jr-sr,i+s}^{(r-1)} \right)\, , \notag 
\end{align}
$\forall \:  k\geq r+1\, , \: i\geq 2\,$. 

Equation~\eqref{hat0_2} can be analyzed similarly as above, by splitting it in three different parts
\begin{align}
\begin{aligned}
\label{secondo_split}
h_{k,0}^{(r)}=&\sum_{j=0}^{\lfloor \frac{k-1}{r} \rfloor -2} \frac{1}{j!}L_{\chi_{2}^{(r)}(\vet{q},\vet{p})}^{j}\left(\widehat{h}_{k-jr,0}^{(r)} \right)  +\frac{1}{\left(\lfloor \frac{k-1}{r} \rfloor -1 \right)!}L_{\chi_{2}^{(r)}(\vet{q},\vet{p})}^{\lfloor (k-1)/r \rfloor -1}\left(\widehat{h}_{k-\left(\lfloor \frac{k-1}{r} \rfloor -1\right) r ,0}^{(r)} \right)\\ 
&+\frac{1}{\lfloor \frac{k-1}{r} \rfloor!}L_{\chi_{2}^{(r)}(\vet{q},\vet{p})}^{\lfloor (k-1)/r \rfloor}\left(\widehat{h}_{k-\lfloor \frac{k-1}{r} \rfloor r,0}^{(r)} \right) 
\end{aligned}
\end{align}
$ \forall \:  k\geq 2r+1\,  $. We study separately the relations satisfied by the indices $(j,k)$ of each of the three terms in the previous equality. Following~\eqref{1gruppo}, we have that
$$k-jr\geq k-\lfloor \frac{k-1}{r} \rfloor r + 2r \geq 2r+1\, ,
$$
(since $j\leq \lfloor (k-1)/r \rfloor -2 \,$). Then, the definition~\eqref{0_4} holds for the first term. 
For the second and third terms, we have different formulas according to whether or not $k$ is a multiple of $r$.
\begin{itemize}
\item[(i)] First case: $k$ is not a multiple of $r\,$, i.e., 
$$k=mr+f\;\; \mathrm{with} \;\; 1\leq f\leq r-1\;\; \mathrm{and}\;\; m\in\mathbb{N}\, .$$
\end{itemize}
From~\eqref{2gruppo_no_multiplo} we have that $1\leq k-\lfloor \frac{k-1}{r} \rfloor r  \leq r-1$ and, consequently, $r+1\leq k-\lfloor \frac{k-1}{r} \rfloor r  + r\leq 2r-1\,$. Then, the definitions~\eqref{0_12} and~\eqref{0_3} hold, respectively, for the third and second term of~\eqref{secondo_split}. Thus, we can write equation~\eqref{secondo_split} as
\begin{align}
h_{k,0}^{(r)}=&\sum_{j=0}^{\lfloor \frac{k-1}{r} \rfloor -2}\sum_{s=0}^{\lfloor \frac{k-jr-1}{r} \rfloor } \frac{1}{j!\,s!}L_{\chi_{2}^{(r)}(\vet{q},\vet{p})}^{j}L_{\chi_{1}^{(r)}(\vet{q})}^{s}\left(h_{k-jr-sr,s}^{(r-1)} \right)  +\frac{1}{\left(\lfloor \frac{k-1}{r} \rfloor -1 \right)!}L_{\chi_{2}^{(r)}(\vet{q},\vet{p})}^{\lfloor (k-1)/r \rfloor -1}\left(h_{k-\left(\lfloor \frac{k-1}{r} \rfloor -1\right) r ,0}^{(r-1)} \right)\, ,\notag
\end{align}
$ \forall \:  k\geq 2r+1\,  $, $k\neq mr\,$, $m\in\mathbb{N}\,$. 
\begin{itemize}
\item[(ii)] Second case: $k$ is a multiple of $r\,$, i.e., 
$$k=mr\;\; \mathrm{with} \;\; m\in\mathbb{N}\, .$$
\end{itemize}
From~\eqref{2gruppo_multiplo} we now have that $k-\lfloor \frac{k-1}{r} \rfloor r =r$ and, consequently, $k-\lfloor \frac{k-1}{r} \rfloor r +r=2r\,$. Then, the definitions~\eqref{0_12} and~\eqref{0_4} hold, respectively, for the third and the second term of~\eqref{secondo_split}. Thus, we can write Eq~\eqref{secondo_split} as
\begin{align}
h_{k,0}^{(r)}=&\sum_{j=0}^{\lfloor \frac{k-1}{r} \rfloor -1}\sum_{s=0}^{\lfloor \frac{k-jr-1}{r} \rfloor } \frac{1}{j!\,s!}L_{\chi_{2}^{(r)}(\vet{q},\vet{p})}^{j}L_{\chi_{1}^{(r)}(\vet{q})}^{s}\left(h_{k-jr-sr,s}^{(r-1)} \right) \, ,
\end{align}
$ \forall \:  k\geq 2r+1\,  $, $k= mr\,$, $m\in\mathbb{N}\,$.
By the same argument, we can write Eq~\eqref{hat1_1NO} as:
\begin{align}
\label{terzo_split}
& h_{k,1}^{(r)}=\sum_{j=0}^{\lfloor \frac{k-1}{r} \rfloor -1} \frac{1}{j!}L_{\chi_{2}^{(r)}(\vet{q},\vet{p})}^{j}\left(\widehat{h}_{k-jr,1}^{(r)} \right) + \frac{1}{\lfloor \frac{k-1}{r} \rfloor!}L_{\chi_{2}^{(r)}(\vet{q},\vet{p})}^{\lfloor (k-1)/r \rfloor}\left(\widehat{h}_{k-\lfloor \frac{k-1}{r} \rfloor r,1}^{(r)} \right) 
\end{align}
$\forall \:  k\geq r+1\, , k\neq m\,r\,  (m\in\mathbb{N})\,$. In view of the inequalities~\eqref{1gruppo} and~\eqref{2gruppo_no_multiplo} we then readily find that the definitions~\eqref{1_2} and ~\eqref{1_0} hold, respectively, for the first and second part of the previous equation, i.e.,~\eqref{terzo_split} leads to 
\begin{align}
h_{k,1}^{(r)}=&\sum_{j=0}^{\lfloor \frac{k-1}{r} \rfloor -1 }\sum_{s=0}^{\lfloor \frac{k-jr-1}{r} \rfloor } \frac{1}{j!\,s!}L_{\chi_{2}^{(r)}(\vet{q},\vet{p})}^{j}L_{\chi_{1}^{(r)}(\vet{q})}^{s}\left(h_{k-jr-sr,1+s}^{(r-1)} \right) \, ,
\end{align}
$\forall \:  k\geq r+1\, , k\neq m\,r\,  (m\in\mathbb{N})\,$. Finally, recalling again~\eqref{1gruppo} and~\eqref{2gruppo_multiplo}, we can write Eq~\eqref{hat1_1SI} as\footnote{In the last equality, due to~\eqref{2gruppo_multiplo}, we have that if $j= \lfloor (k-1)/r \rfloor\,$, then $\lfloor(k-j r -1)/r \rfloor=\lfloor(k-\lfloor(k-1)/r \rfloor r -1)/r \rfloor =\lfloor (r-1)/r \rfloor = 0\, $, i.e., $s=0\,$. Thus, also here we can join the terms in a single sum.}
\begin{align}
\begin{aligned}
 h_{k,1}^{(r)}=&\sum_{j=0}^{\lfloor \frac{k-1}{r} \rfloor -1} \frac{1}{j!}L_{\chi_{2}^{(r)}(\vet{q},\vet{p})}^{j}\left(\widehat{h}_{k-jr,1}^{(r)} \right) + \frac{1}{\lfloor \frac{k-1}{r} \rfloor!}L_{\chi_{2}^{(r)}(\vet{q},\vet{p})}^{\lfloor (k-1)/r \rfloor}\left(\widehat{h}_{k-\lfloor \frac{k-1}{r} \rfloor r,1}^{(r)} \right)+ \frac{1}{m!} L_{\chi_{2}^{(r)}(\vet{q},\vet{p})}^{m}\left(\vet{\omega}\cdot\vet{p}\right)\\
\phantom{h_{k,1}^{(r)}}=& \sum_{j=0}^{\lfloor \frac{k-1}{r} \rfloor -1}\sum_{s=0}^{\lfloor \frac{k-jr-1}{r} \rfloor } \frac{1}{j!\,s!}L_{\chi_{2}^{(r)}(\vet{q},\vet{p})}^{j}L_{\chi_{1}^{(r)}(\vet{q})}^{s}\left(h_{k-jr-sr,1+s}^{(r-1)} \right)  + \frac{1}{\lfloor \frac{k-1}{r} \rfloor!}L_{\chi_{2}^{(r)}(\vet{q},\vet{p})}^{\lfloor (k-1)/r \rfloor}\left(h_{k-\lfloor \frac{k-1}{r} \rfloor r,1}^{(r-1)} \right)\\
&+ \frac{1}{m!} L_{\chi_{2}^{(r)}(\vet{q},\vet{p})}^{m}\left(\vet{\omega}\cdot\vet{p}\right)= \\
=& \sum_{j=0}^{\lfloor \frac{k-1}{r} \rfloor }\sum_{s=0}^{\lfloor \frac{k-jr-1}{r} \rfloor } \frac{1}{j!\,s!}L_{\chi_{2}^{(r)}(\vet{q},\vet{p})}^{j}L_{\chi_{1}^{(r)}(\vet{q})}^{s}\left(h_{k-jr-sr,1+s}^{(r-1)} \right) + \frac{1}{m!} L_{\chi_{2}^{(r)}(\vet{q},\vet{p})}^{m}\left(\vet{\omega}\cdot\vet{p}\right)\, ,
\end{aligned}
\end{align}
$\forall \:  k\geq r+1\, , k= m\,r\,\,  (m\in\mathbb{N})\,$ (where we used the definitions~\eqref{1_2} and~\eqref{1_1} for the first and second part of the sum). As before, we can then write the previous equation as
\begin{equation}
\begin{split}
&\sum_{j=0}^{m-1 }\sum_{s=0}^{\lfloor \frac{mr-jr-1}{r} \rfloor } \frac{1}{j!\,s!}L_{\chi_{2}^{(r)}(\vet{q},\vet{p})}^{j}L_{\chi_{1}^{(r)}(\vet{q})}^{s}\left(h_{mr-jr-sr,1+s}^{(r-1)} \right) + \frac{1}{m!} L_{\chi_{2}^{(r)}(\vet{q},\vet{p})}^{m}\left(\vet{\omega}\cdot\vet{p}\right)=\\
&\sum_{j=0}^{m-2 }\sum_{s=0}^{\lfloor \frac{mr-jr-1}{r} \rfloor } \frac{1}{j!\,s!}L_{\chi_{2}^{(r)}(\vet{q},\vet{p})}^{j}L_{\chi_{1}^{(r)}(\vet{q})}^{s}\left(h_{mr-jr-sr,1+s}^{(r-1)} \right)+\frac{1}{(m-1)!}L_{\chi_{2}^{(r)}(\vet{q},\vet{p})}^{m-1}\left(h_{r,1}^{(r-1)} \right) + \frac{1}{m!} L_{\chi_{2}^{(r)}(\vet{q},\vet{p})}^{m}\left(\vet{\omega}\cdot\vet{p}\right)=\\
&\sum_{j=0}^{m-2 }\sum_{s=0}^{\lfloor \frac{mr-jr-1}{r} \rfloor } \frac{1}{j!\,s!}L_{\chi_{2}^{(r)}(\vet{q},\vet{p})}^{j}L_{\chi_{1}^{(r)}(\vet{q})}^{s}\left(h_{mr-jr-sr,1+s}^{(r-1)} \right)+\frac{1}{m!}L_{\chi_{2}^{(r)}(\vet{q},\vet{p})}^{m-1}\left(h_{r,1}^{(r-1)}+ L_{\chi_{2}^{(r)}(\vet{q},\vet{p})}\left(\vet{\omega}\cdot\vet{p}\right)\right) \\
&+\frac{m-1}{m!}L_{\chi_{2}^{(r)}(\vet{q},\vet{p})}^{m-1}\left(h_{r,1}^{(r-1)}\right)\, . 
\end{split}
\end{equation}
Use of Eq~\eqref{1_1} and the homological equation~\eqref{uccidi.term.p} then concludes the proof.\qed

\section{Concluding remarks}
\label{Conclusion}
The focus of the present paper is twofold.

i) We emphasize the differences between two distinct methods by which Lindstedt series can be
computed in nonlinearly coupled oscillator Hamiltonian models. In particular, we discussed a
Lindstedt method called `analogous to the Birkhoff series', and another called `analogous to
the Kolmogorov series'. In the first case, series expansions are defined in open domains
in the action space around the origin, and the frequencies are represented as (series) polynomial
functions of the action variables. In the second method, instead, the frequencies along any
required torus solution must be fixed in advance (`torus fixing'), and the series allow (after
reversion) to determine a posteriori the amplitudes (or values of the actions) for which
the motion takes place with a given set of frequencies. We explain how this difference in
the physical interpretation of what the series are meant to compute leads also to formal
differences in the way the series terms are computed, as well as to real differences in
the convergence and precision properties of the series.

ii) We propose and give the formal structure of an algorithm viewed as an analogue of
the Kolmogorov scheme within the framework of the `torus fixing' method, but properly dealing
with the lack of the twist property in the Hamiltonian appearing in the kernel of the
associated perturbative scheme. In particular, we demonstrate how the lack of the twist
condition can be compensated by a suitable introduction of counter-terms (in the spirit
of the procedures proposed already in~\cite{bargen2002,gal1994a,gal1994b,sandan2021}). In fact, we demonstrate how to compute
such counter-terms in a way leading to precisely the same frequency corrections as those
obtained by the Lindstedt method called `analogous to Kolmogorov'.

Our formal algorithm given in Section 3 above is accompanied by particular examples
offering some intuition into the subtleties of each examined method. On the other hand, as stressed in the introduction, the convergence of the Kolmogorov series computed by the algorithm presented in Section~\ref{KAM} can only be inferred by an indirect argument, namely, their equivalence with the Lindstedt series of Subsection~\ref{Lind.classico.even}. Thus, we point out the interesting open question of a direct proof of the convergence of our hereby presented KAM algorithm for isochronous systems (see also~\cite{sandan2021}).


\appendix
\section{Example -- 1DOF Hamiltonian with an odd (cubic) degree dependence on $x$}
\label{appendixA}
Consider the following one-degree of freedom Hamiltonian, with odd power dependence on the variable $x$
\begin{equation}
\label{Ham.es.odd}
\Hscr(x, p)=\frac{\omega_0}{2}\left(p^2+x^2\right)-\varepsilon \frac{x^3}{3}\,.
\end{equation}
We pass to action-angle variables $(J,\,q)$ through the transformation $x=\sqrt{2J}\sin(q)\,$, $p=\sqrt{2 J}\cos(q)\,$, obtaining
\begin{equation}
\label{Ham.es.Jq.odd}
\Hscr(q,J)= \omega_0 \,J  - \frac{\varepsilon}{\sqrt{2}}  J^{3/2} \sin(q) + \frac{\varepsilon}{3\,\sqrt{2}}  J^{3/2} \sin(3 q)\, .
\end{equation}
Finally, we define the translation $J=J_0+p\,$ leading to (apart from constants)
\begin{equation}
\label{Ham.es.pq.odd}
\begin{split}
\Hscr(q,p)= \omega_0\,p  - \frac{\varepsilon}{\sqrt{2}}  (J_0+p)^{3/2} \sin(q) + \frac{\varepsilon}{3\,\sqrt{2}}  (J_0+p)^{3/2} \sin(3 q)\, .
\end{split}
\end{equation}
Since the variable $J=J_0+p$ appears in the above Hamiltonian in half-integer powers, we expand the Hamiltonian in powers of the variable $p\,$ up to the same order as the maximum normalization order in $\varepsilon$. This proves to be sufficient since higher powers of $p$ only influence the process at powers (in the book keeping order) higher than the maximum normalization order. In particular, the following lemma can be easily proved:
\newtheorem{cnt.lemma}{Lemma}
\setcounter{cnt.lemma}{0}
\label{lemma1}
\begin{cnt.lemma}
Let
$\Hscr(q, J)=\omega_0\,J + \varepsilon\,h(q,J)\,$, where $h=\Oscr(J^\frac{k}{2})\,$ $k\geq 3\,$. Then, for all $n\geq 1$ (with $n=$ number of normalization steps) $\chi_1^{(n)}=\Oscr\left(J_0^{\frac{(k-2)n+2}{2}}\right)$ and $\chi_2^{(n)}=\Oscr\left(J_0^{\frac{(k-2)n}{2}}\right)\,$.
\end{cnt.lemma}
\begin{proof}
For the proof, see the appendix~\ref{appendixB}.
\end{proof}
We will now illustrate the method by computing the direct (Lindstedt) and indirect (Kolmogorov) series up to order $3$ in $\varepsilon\,$ in the example above (the reason for reaching order $3$ instead of $2$ will become clear below). Starting from the Hamiltonian~\eqref{Ham.es.pq.odd} we perform an expansion in $p$ leading to (apart from constants)
\begin{equation}
\label{Ham.iniz.odd}
\begin{split}
\Hscr(q,p)&=\omega_0\,p-\frac{\varepsilon\,J_0^{3/2} }{\sqrt{2}}\,\sin (q)-\frac{3 \,\varepsilon\, \sqrt{J_0} }{2 \,\sqrt{2}}\,p\, \sin (q)-\frac{3 \,\varepsilon}{8 \,\sqrt{2} \,\sqrt{J_0}}\, p^2 \,\sin (q)+\frac{\varepsilon }{16 \,\sqrt{2} \,J_0^{3/2}}\,p^3 \,\sin (q)\\
&\phantom{=}+\frac{\varepsilon\,J_0^{3/2} }{3\, \sqrt{2}}\,\sin (3 q)+\frac{\varepsilon\, \sqrt{J_0} }{2\, \sqrt{2}}\,p \,\sin (3 q)+\frac{\varepsilon}{8 \,\sqrt{2}\, \sqrt{J_0}}\, p^2\, \sin (3 q)-\frac{\varepsilon }{48\, \sqrt{2} \,J_0^{3/2}}\,p^3\, \sin (3 q)\, .
\end{split}
\end{equation}  

\subsection{Lindstedt series analogous to Kolmogorov}
Following the same procedure as in subsection~\ref{Lind.classico.even}, we start with the equations of motion under the Hamiltonian~\eqref{Ham.es.Jq.odd}
\begin{subnumcases}{}
\label{eq.moto.odd.LK.q}
&$\displaystyle{ \dot{q}(t)=\frac{\partial \Hscr}{\partial J}=
\omega_0-\frac{3\,\varepsilon\, \sqrt{J} }{2\, \sqrt{2}}\,\sin (q)+\frac{\varepsilon\, \sqrt{J} }{2 \,\sqrt{2}}\,\sin (3 q)}$\\
\label{eq.moto.odd.LK.J}
&$\displaystyle{\dot{J}(t)=-\frac{\partial \Hscr}{\partial q}=
\frac{\varepsilon\, J^{3/2} }{\sqrt{2}}\,\cos (q)-\frac{\varepsilon\, J^{3/2}}{\sqrt{2}}\, \cos (3 q)}\,.$
\end{subnumcases}
Replacing, as before, the expressions
\begin{equation}
\begin{split}
&\omega_0=\omega+\varepsilon\,a_1+\varepsilon^2\,a_2+\varepsilon^3\,a_3\, , \\
&q(t)=q_0(t)+\varepsilon\,q_1(t)+\varepsilon^2\,q_2(t)+\varepsilon^3\,q_3(t)\, , \\
&J(t)=J_0(t)+\varepsilon J_1(t)+\varepsilon^2 J_2(t)+\varepsilon^3 J_3(t)\, ,
\end{split}
\end{equation}
into the equations of motion and performing an expansion up to order $3$ in $\varepsilon\,$ (having fixed $\omega$) we compare terms of like orders in $\varepsilon$ in the l.h.s and r.h.s of~\eqref{eq.moto.odd.LK.q} and~\eqref{eq.moto.odd.LK.J}. At order zero we have
$$
\dot{q_0}(t)=\omega \, ,\qquad\dot{J_{0}}(t)=0\quad\Longrightarrow\quad q_0(t)=\omega\,t\, , \qquad J_0(t)=J_0\, ,
$$
where we fix the initial phase $q_0(0)=0$ and $J_0(0)=J_0\,$. At order one, we find 
\begin{equation}
\begin{split}
&\displaystyle{ \dot{q_{1}}(t)=
 a_1-\frac{3 \,\sqrt{J_0}}{2\, \sqrt{2}}\, \sin (\omega\,t)+\frac{\sqrt{J_0} }{2 \,\sqrt{2}}\,\sin (3 \,\omega\,t) },\\
&\displaystyle{\dot{J_{1}}(t)=
\frac{J_0^{3/2} }{\sqrt{2}}\,\cos (\omega\,t)-\frac{J_0^{3/2} }{\sqrt{2}}\,\cos (3 \,\omega\,t)}
\, .
\end{split}
\end{equation}
Since no constant terms arise in $\dot{q_1}(t)$, we have that $a_1=0$. Then
$$
q_1(t)=-\frac{2\, \sqrt{2}\, \sqrt{J_0}}{3 \,\omega}+\frac{3\, \sqrt{J_0}}{2 \,\sqrt{2}\,\omega}\, \cos (\omega\,t)-\frac{\sqrt{J_0} }{6 \,\sqrt{2}\,\omega}\,\cos (3\,\omega\,t)\, ,\quad J_1(t)=\frac{J_0^{3/2}}{\sqrt{2} \,\omega}\, \sin (\omega\,t)-\frac{J_0^{3/2} }{3 \,\sqrt{2} \,\omega}\,\sin (3 \,\omega\,t)
$$
yielding the constants $q_1(0)=J_1(0)=0\,$. At order two we now get 
\begin{equation}
\begin{split}
&\displaystyle{ \dot{q_{2}}(t)=a_2-\frac{5 \,J_0}{6 \,\omega}+\frac{J_0}{\omega}\, \cos (\omega\,t)+\frac{3 \,J_0}{8 \,\omega}\, \cos (2\,\omega\,t)-\frac{J_0}{\omega}\, \cos (3\,\omega\,t)+\frac{J_0}{2 \,\omega}\, \cos (4\,\omega\,t)-\frac{J_0}{24 \,\omega}\, \cos (6\,\omega\,t)} ,\\
&\displaystyle{\dot{J_{2}}(t)=\frac{2 \,J_0^2}{3 \,\omega}\, \sin (\omega\,t)+\frac{4 \,J_0^2 }{3 \,\omega}\,\sin (2\,\omega\,t)-\frac{2\,J_0^2 }{\omega}\,\sin (3 \,\omega\,t)+\frac{2 \,J_0^2}{3 \,\omega}\, \sin (4\,\omega\,t)
}
\, ;
\end{split}
\end{equation}
To compensate for the constant term in $\dot{q_2}(t)\,$ we now set $a_2=5\,J_0/(6\,\omega)$. Then, 
\begin{equation}
\begin{split}
&\displaystyle{ q_{2}(t)=\frac{J_0\, }{\omega^2}\,\sin (\omega\,t)+\frac{3\,J_0}{16\,\omega^2}\, \sin (2\,\omega\,t)-\frac{J_0}{3 \,\omega^2}\, \sin (3 \,\omega\,t)+\frac{J_0}{8 \,\omega^2}\, \sin (4 \,\omega\,t)-\frac{J_0}{144 \,\omega^2}\,\sin (6 \,\omega\,t)} ,\\
&\displaystyle{J_{2}(t)=\frac{5 \,J_0^2}{6 \,\omega^2}-\frac{2\, J_0^2 }{3\,\omega^2}\,\cos (\omega\,t)-\frac{2\,J_0^2 }{3 \,\omega^2}\,\cos (2\,\omega\,t)+\frac{2 \,J_0^2 }{3 \,\omega^2}\,\cos (3\,\omega\,t)-\frac{J_0^2}{6 \,\omega^2}\, \cos (4\,\omega\,t)
}
\, .
\end{split}
\end{equation}
In a similar way, at order $3$ we find $a_3=0$, and the solutions
\begin{equation}
\label{sol.Lind.odd}
\begin{split}
&q(t)= \omega\,t-\frac{2
\,   \sqrt{2} \,\varepsilon\, \sqrt{J_0}}{3 \,\omega}+\frac{3 \,\varepsilon\, \sqrt{J_0} }{2 \,\sqrt{2} \,\omega}\,\cos (\omega\,t)-\frac{\varepsilon \,\sqrt{J_0} }{6 	\,\sqrt{2}\,\omega}\,\cos (3 \,\omega\,t)+\frac{\,\varepsilon^2\, J_0}{\omega^2}\, \sin (\omega\,t)+\frac{3 \,\varepsilon^2\, J_0}{16\, \omega^2}\, \sin (2 \,\omega\,t)\\
&\phantom{q(t)=}-\frac{\varepsilon^2\, J_0
  }{3 \,\omega^2}\, \sin (3 \,\omega\,t)+\frac{\varepsilon^2\, J_0}{8 \,\omega^2}\, \sin (4 \,\omega\,t)-\frac{\varepsilon^2\, J_0 }{144 \,\omega^2}\,\sin (6 \,\omega\,t)-\frac{38\, \sqrt{2} \,\varepsilon^3\, J_0^{3/2}}{81 \omega^3}+\frac{\varepsilon^3 \,J_0^{3/2} }{\sqrt{2}\, \omega^3}\,\cos (\omega\,t)\\
&\phantom{q(t)=}-\frac{\varepsilon^3\, J_0^{3/2} }{2 \,\sqrt{2}\,\omega^3}\,\cos (2 \,\omega\,t)+\frac{145 \,\varepsilon^3\, J_0^{3/2} }{144\, \sqrt{2}\, \omega^3}\,\cos (3 \,\omega\,t)-\frac{\sqrt{2} \,\varepsilon^3\, J_0^{3/2} }{3\, \omega^3}\,\cos (4
   \,\omega\,t)+\frac{\varepsilon^3\, J_0^{3/2}}{16 \,\sqrt{2} \,\omega^3}\, \cos (5 \,\omega\,t)\\
&\phantom{q(t)=}+\frac{\varepsilon^3\, J_0^{3/2}}{18\, \sqrt{2} \,\omega^3}\, \cos (6 \,\omega\,t)-\frac{\varepsilon^3\, J_0^{3/2} }{48 \,\sqrt{2}\, \omega^3}\,\cos (7 \,\omega\,t)+\frac{\varepsilon^3\, J_0^{3/2} }{1296\, \sqrt{2}\, \omega^3}\,\cos(9 \,\omega\,t)\, , \\[1.3ex]
&J(t)=J_0+\frac{\varepsilon\, J_0^{3/2}}{\sqrt{2} \,\omega}\, \sin (\omega\,t)-\frac{\varepsilon\, J_0^{3/2} }{3\, \sqrt{2}\, \omega}\,\sin (3 \,\omega\,t)+\frac{5 \,\varepsilon^2\, J_0^2}{6\, \omega^2}-\frac{2 \,\varepsilon^2 \,J_0^2 }{3\, \omega^2}\,\cos (\omega\,t)-\frac{2 \,\varepsilon^2\, J_0^2 }{3\, \omega^2}\,\cos (2 \,\omega\,t)\\
&\phantom{J(t)=}+\frac{2 \,\varepsilon^2\,
   J_0^2 }{3 \,\omega^2}\,\cos (3 \,\omega\,t)-\frac{\varepsilon^2\, J_0^2 }{6\, \omega^2}\,\cos (4 \,\omega\,t)+\frac{7 \,\varepsilon^3\, J_0^{5/2} }{4 \,\sqrt{2}\, \omega^3}\,\sin (\omega\,t)-\frac{8\, \sqrt{2} \,\varepsilon^3\, J_0^{5/2} }{9\,\omega^3}\,\sin (2 \,\omega\,t)\\
  &\phantom{J(t)=}+\frac{13 \,\varepsilon^3\, J_0^{5/2} }{8 \,\sqrt{2}\, \omega^3}\,\sin (3 \,\omega\,t)-\frac{4\, \sqrt{2} \,\varepsilon^3\, J_0^{5/2}}{9\, \omega^3}\, \sin (4 \,\omega\,t)+\frac{7 \,\varepsilon^3\, J_0^{5/2} }{72\, \sqrt{2}\, \omega^3}\,\sin (5 \,\omega\,t)\, ,\\[1.3ex]
&\omega=\omega_0-\frac{5 \,\varepsilon^2 J_0}{6\, \omega}\, .
\end{split}
\end{equation}

\subsection{Kolmogorov normal form} 
Starting from the Hamiltonian~\eqref{Ham.iniz.odd}, we perform canonical transformations in order to bring the Hamiltonian into Kolmogorov normal form, i.e., $\Hscr(q,p)=\omega\,p+\varepsilon\,R_1(q,\,p)+\varepsilon^2\,R_2(q,\,p)+\varepsilon^3\,R_3(q,\,p)$ where $R_{i}(q,\,p)=\Oscr(\norm{p}^2)$ $i=1,\,2,\,3$. Substituting, as in Section~\ref{sec2}, the expression $\omega_0=\omega + \varepsilon\,a_1+\varepsilon^2\,a_2+\varepsilon^3\,a_3\,$ in the Hamiltonian~\eqref{Ham.iniz.odd}, we have
\begin{equation}
\label{Ham.iniz.odd.KAM}
\begin{split}
\Hscr(q,p)&=\omega\,p + \varepsilon\,a_1\,p-\frac{\varepsilon\,J_0^{3/2} }{\sqrt{2}}\,\sin (q)-\frac{3 \,\varepsilon\, \sqrt{J_0} }{2 \,\sqrt{2}}\,p\, \sin (q)-\frac{3 \,\varepsilon}{8 \,\sqrt{2} \,\sqrt{J_0}}\, p^2 \,\sin (q)\\
&\phantom{=}+\frac{\varepsilon }{16 \,\sqrt{2} \,J_0^{3/2}}\,p^3 \,\sin (q)+\frac{\varepsilon\,J_0^{3/2} }{3\, \sqrt{2}}\,\sin (3 q)+\frac{\varepsilon\, \sqrt{J_0} }{2\, \sqrt{2}}\,p \,\sin (3 q)+\frac{\varepsilon}{8 \,\sqrt{2}\, \sqrt{J_0}}\, p^2\, \sin (3 q)\\
&\phantom{=}-\frac{\varepsilon }{48\, \sqrt{2} \,J_0^{3/2}}\,p^3\, \sin (3 q)+\varepsilon^2\,a_2\,p +\varepsilon^3\,a_3\,p\, .
\end{split}
\end{equation} 
At first order, we have
\begin{equation}
\label{a1.K.odd}
\avg{h_{1,1}^{(0)}}_{{q}}=\avg{ a_1\, p-\frac{3 \, \sqrt{J_0}}{2 \,\sqrt{2}}\, p\, \sin (q)+\frac{\sqrt{J_0}}{2\, \sqrt{2}}\, p\, \sin (3\, q)}_{{q}}=0\qquad\Longrightarrow\qquad a_1=0\, ; 
\end{equation}
implying that $a_1=0$ for the corresponding counterterm in the Hamiltonian~\eqref{Ham.iniz.odd.KAM}. In order to eliminate the terms constant in the actions (depending only in the angle $q\,$) $\varepsilon\,h_{1,0}$, given by
$$
h_{1,0}^{(0)}(q)=-\frac{J_0^{3/2} }{\sqrt{2}}\sin (q)+\frac{J_0^{3/2} }{3 \,\sqrt{2}}\,\sin (3\, q)\, ,
$$
we define the generating function $X^{(1)}(q)$ such that $\displaystyle{L_{X^{(1)}(q)}(\omega\,p)+h_{1,0}^{(0)}=\avg{h_{1,0}^{(0)}}_{{q}}}\,$. Hence
$$
X^{(1)}=\frac{J_0^{3/2} }{\sqrt{2} \,\omega}\,\cos (q)-\frac{J_0^{3/2}}{9 \,\sqrt{2} \,\omega}\, \cos (3\, q)\, .
$$
In order to fix the initial value of $p$ at zero, we then set (as in Section~\ref{sec2}) $\chi_{1}^{(1)}(q)=X^{(1)}(q) + K^{(1)}\,q\,$; in this case, the constant $K^{(1)}=0\,$.
We can now determine the intermediate Hamiltonian 
$$ 
\widehat{\Hscr}^{(1)}=\exp\left(L_{\varepsilon\,\chi_1^{(1)}(q)}\right)\Hscr\, . 
$$

Denoting by $\varepsilon\,\widehat{h}_{1,1}^{(1)}$ the terms of $\widehat{\Hscr}^{(1)}$ of order one in $\varepsilon$ and linear in $p\,$, we have 
$$\widehat{h}_{1,1}^{(1)}=-\frac{3 \, \sqrt{J_0}\, p}{2\, \sqrt{2}}\, \sin (q)+\frac{ \sqrt{J_0} \,p }{2 \,\sqrt{2}}\,\sin (3\, q)\, .
$$
These terms are eliminated by the generating function $\chi_{2}^{(1)}(q,\,p)=\tilde{\chi}_{2}^{(1)}(q,\,p)+S^{(1)}\,p$ satisfying the homological equation $L_{\chi_{2}^{(1)}(q,\,p)}(\omega\,p)+\widehat{h}_{1,1}^{(1)}=0\,$. In order to fix the initial value of $q$ at zero, we readily find 
$$
S^{(1)}=-\frac{2\, \sqrt{2}\, \sqrt{J_0} }{3 \,\omega} 
$$ 
and
$$\chi_{2}^{(1)}(q,\,p)=-\frac{2\, \sqrt{2} \, \sqrt{J_0} \,p}{3 \,\omega}+\frac{3\, \sqrt{J_0} \,p }{2 \,\sqrt{2}\,\omega}\,\cos (q)-\frac{\sqrt{J_0}\, p}{6 \,\sqrt{2} \,\omega}\, \cos (3\, q)\, .
$$
Thus, the Hamiltonian 
$$ 
\Hscr^{(1)}=\exp\left(L_{\varepsilon\,\chi_{2}^{(1)}(q,\,p)}\right)\widehat{\Hscr}^{(1)}\, 
$$
is now in Kolmogorov normal form up to first order in $\varepsilon\,$ (it is easy to check that the transformed Hamiltonian contains the normal form terms $-\displaystyle{\frac{3\, p^2 }{8\, \sqrt{2} \,\sqrt{J_0}}\,\sin (q)}+\displaystyle{\frac{p^2}{8 \,\sqrt{2} \,\sqrt{J_0}}\, \sin (3\, q)}+\displaystyle{\frac{p^3 }{16\, \sqrt{2}\, J_0^{3/2}}\,\sin (q)}-\displaystyle{\frac{p^3}{48\, \sqrt{2} \,J_0^{3/2}}\, \sin (3\, q)}\, $). 

In a similar way we can proceed at orders 2 and 3, obtaining the formulas:
\begin{equation}
\label{a2.K.odd}
\begin{split}
&a_2=\frac{5 \,J_0}{6\,\omega}\, , \\
&\chi_{1}^{(2)}=-\frac{5 \,J_0^2\, q}{12\,\omega^2}+\frac{5\,J_0^2 }{16\,\omega^2}\,\sin (2\, q)-\frac{J_0^2}{16 \,\omega^2}\, \sin (4\, q)+\frac{J_0^2}{144 \,\omega^2}\, \sin (6\, q)\, ,  \\
&\chi_{2}^{(2)}=\frac{J_0\, p }{2 \,\omega^2}\,\sin (q)+\frac{13 \,J_0 \,p }{32 \,\omega^2}\,\sin (2\, q)-\frac{J_0\,   p }{6 \,\omega^2}\,\sin (3\, q)+\frac{J_0\, p}{288\, \omega^2}\, \sin (6\, q) 
\end{split}
\end{equation}
and 
\begin{align}
\label{a3.K.odd}
\begin{split}
&a_3=0\, , \\
&\chi_{1}^{(3)}=\frac{49\,J_0^{5/2} }{96 \,\sqrt{2}\, \omega^3}\,\cos (q)-\frac{5 \,J_0^{5/2} }{6 \,\sqrt{2}\,\omega^3}\,\cos (2\, q)+\frac{43\, J_0^{5/2} }{432\, \sqrt{2} \,\omega^3}\,\cos (3\, q)+\frac{J_0^{5/2} }{3\,\sqrt{2}\,\omega^3}\,\cos (4\, q)\\
&\phantom{\chi_{1}^{(3)}=}-\frac{29 \,J_0^{5/2}}{240\, \sqrt{2}\, \omega^3}\, \cos (5\, q)-\frac{J_0^{5/2} }{18\, \sqrt{2}\, \omega^3}\,\cos (6\, q)+\frac{7 \,J_0^{5/2} }{192 \,\sqrt{2} \,\omega^3}\,\cos (7\, q)-\frac{11\,J_0^{5/2}}{5184\, \sqrt{2}\, \omega^3}\, \cos (9\, q)\, ,  \\
&\chi_{2}^{(3)}=-\frac{107\,J_0^{3/2} \,p}{162\, \sqrt{2} \,\omega^3}+\frac{295\,J_0^{3/2}\, p }{192\, \sqrt{2}\,\omega^3}\,\cos (q)-\frac{47 \,J_0^{3/2} \,p }{36\, \sqrt{2} \,\omega^3}\,\cos (2 \,q)+\frac{133 \,J_0^{3/2} \,p }{288\, \sqrt{2}\,\omega^3}\,\cos (3 \,q) \\
&\phantom{\chi_{2}^{(3)}=}-\frac{J_0^{3/2} \,p }{18\, \sqrt{2}\, \omega^3}\,\cos(4 \,q)+\frac{13\,J_0^{3/2}\, p }{288 \,\sqrt{2}\, \omega^3}\,\cos (5\, q)-\frac{J_0^{3/2}\, p }{36 \,\sqrt{2}\,\omega^3}\,\cos (6\, q)+\frac{7 \,J_0^{3/2} \,p}{1152\, \sqrt{2}\,\omega^3}\, \cos (7\, q) \\
&\phantom{\chi_{2}^{(3)}=}-\frac{J_0^{3/2}\, p }{10368 \,\sqrt{2} \,\omega^3}\,\cos (9\, q) \, ,
\end{split}
\end{align}
for the generating functions. The final Hamiltonian is 
$$\Hscr^{(3)}(\q,\p)=\omega\,\p+ \varepsilon\,R_1(q,\,p)+\varepsilon^2\,R_2(q,\,p)+\varepsilon^3\,R_3(q,\,p)
$$
where $(\q,\,\p)$ indicate the new variables, and $R_j(\q,\p)=\Oscr(\norm{\p}^2)$ $j=1,2,3\,$. Thus, the torus $\p(t)=0,\,\q(t)=\omega\,t\,$ is a solution for the equations of motion of this Hamiltonian. Using the Lie transformations, the solution in the original variables reads
\begin{align}
&q(t)= \omega\,t-\frac{2
\,   \sqrt{2} \,\varepsilon\, \sqrt{J_0}}{3 \,\omega}+\frac{3 \,\varepsilon\, \sqrt{J_0} }{2 \,\sqrt{2} \,\omega}\,\cos (\omega\,t)-\frac{\varepsilon \,\sqrt{J_0} }{6 	\,\sqrt{2}\,\omega}\,\cos (3 \,\omega\,t)+\frac{\,\varepsilon^2\, J_0}{\omega^2}\, \sin (\omega\,t)+\frac{3 \,\varepsilon^2\, J_0}{16\, \omega^2}\, \sin (2 \,\omega\,t)\notag\\
&\phantom{q(t)=}-\frac{\varepsilon^2\, J_0
  }{3 \,\omega^2}\, \sin (3 \,\omega\,t)+\frac{\varepsilon^2\, J_0}{8 \,\omega^2}\, \sin (4 \,\omega\,t)-\frac{\varepsilon^2\, J_0 }{144 \,\omega^2}\,\sin (6 \,\omega\,t)-\frac{38\, \sqrt{2} \,\varepsilon^3\, J_0^{3/2}}{81 \omega^3}+\frac{\varepsilon^3 \,J_0^{3/2} }{\sqrt{2}\, \omega^3}\,\cos (\omega\,t)\notag\\
&\phantom{q(t)=}-\frac{\varepsilon^3\, J_0^{3/2} }{2 \,\sqrt{2}\,\omega^3}\,\cos (2 \,\omega\,t)+\frac{145 \,\varepsilon^3\, J_0^{3/2} }{144\, \sqrt{2}\, \omega^3}\,\cos (3 \,\omega\,t)-\frac{\sqrt{2} \,\varepsilon^3\, J_0^{3/2} }{3\, \omega^3}\,\cos (4   \,\omega\,t)+\frac{\varepsilon^3\, J_0^{3/2}}{16 \,\sqrt{2} \,\omega^3}\, \cos (5 \,\omega\,t)\notag\\
&\phantom{q(t)=}+\frac{\varepsilon^3\, J_0^{3/2}}{18\, \sqrt{2} \,\omega^3}\, \cos (6 \,\omega\,t)-\frac{\varepsilon^3\, J_0^{3/2} }{48 \,\sqrt{2}\, \omega^3}\,\cos (7 \,\omega\,t)+\frac{\varepsilon^3\, J_0^{3/2} }{1296\, \sqrt{2}\, \omega^3}\,\cos(9 \,\omega\,t)\, , \notag\\[1.3ex]
&p(t)=\frac{\varepsilon\, J_0^{3/2}}{\sqrt{2} \,\omega}\, \sin (\omega\,t)-\frac{\varepsilon\, J_0^{3/2} }{3\, \sqrt{2}\, \omega}\,\sin (3 \,\omega\,t)+\frac{5 \,\varepsilon^2\, J_0^2}{6\, \omega^2}-\frac{2 \,\varepsilon^2 \,J_0^2 }{3\, \omega^2}\,\cos (\omega\,t)-\frac{2 \,\varepsilon^2\, J_0^2 }{3\, \omega^2}\,\cos (2 \,\omega\,t)\notag\\
&\phantom{J(t)=}+\frac{2 \,\varepsilon^2\,
   J_0^2 }{3 \,\omega^2}\,\cos (3 \,\omega\,t)-\frac{\varepsilon^2\, J_0^2 }{6\, \omega^2}\,\cos (4 \,\omega\,t)+\frac{7 \,\varepsilon^3\, J_0^{5/2} }{4 \,\sqrt{2}\, \omega^3}\,\sin (\omega\,t)-\frac{8\, \sqrt{2} \,\varepsilon^3\, J_0^{5/2} }{9\,\omega^3}\,\sin (2 \,\omega\,t)\notag\\
  &\phantom{J(t)=}+\frac{13 \,\varepsilon^3\, J_0^{5/2} }{8 \,\sqrt{2}\, \omega^3}\,\sin (3 \,\omega\,t)-\frac{4\, \sqrt{2} \,\varepsilon^3\, J_0^{5/2}}{9\, \omega^3}\, \sin (4 \,\omega\,t)+\frac{7 \,\varepsilon^3\, J_0^{5/2} }{72\, \sqrt{2}\, \omega^3}\,\sin (5 \,\omega\,t)\, ,\notag\\
  &J(t)=J_0+p(t)\, .
\end{align}
Recalling also the computed values of $a_1\,$, $a_2\,$ and $a_3$ (Eqs~\eqref{a1.K.odd}--\eqref{a3.K.odd}) we have also
$$\omega=\omega_0-\frac{5 \,\varepsilon^2 J_0}{6\, \omega}\, .
$$
Thus, we obtain the same solutions as by the Lindstedt method (Eq~\eqref{sol.Lind.odd}). 

We remark that also in this case the solutions produced by the Birkhoff normal form are equal to those produced by the Lindstedt method in the version `analogous to Birkhoff'. However, if we compare these solutions with those produced by the KAM algorithm, we also note many differences from order $3$ and beyond. For completeness, we report in the following the solutions with the Birkhoff method:  
\begin{align}
&q(t)=\omega\,t -\frac{2 \,\sqrt{2}\, \varepsilon\, \sqrt{J_0}}{3 \omega_0}+\frac{3\, \varepsilon\, \sqrt{J_0} }{2\, \sqrt{2}\, \omega_0}\,\cos (\omega\,t)-\frac{\varepsilon\, \sqrt{J_0}}{6 \,\sqrt{2}\,   \omega_0}\, \cos (3\, \omega\,t)+\frac{\varepsilon^2\, J_0}{\omega_0^2}\,\sin (\omega\,t)+\frac{3 \,\varepsilon^2\, J_0}{16\,   \omega_0^2}\, \sin (2\, \omega\,t)\notag\\
&\phantom{q(t)=}-\frac{\varepsilon^2\, J_0}{3\, \omega_0^2}\, \sin (3 \,\omega\,t)+\frac{\varepsilon^2\, J_0 }{8\,
   \omega_0^2}\,\sin (4 \,\omega\,t)-\frac{\varepsilon^2\, J_0}{144\, \omega_0^2}\, \sin (6\, \omega\,t)-\frac{83\, \sqrt{2}\,  \varepsilon^3\, J_0^{3/2}}{81\, \omega_0^3}+\frac{9\, \varepsilon^3 \,J_0^{3/2} }{4 \,\sqrt{2}\, \omega_0^3}\,\cos (\omega\,t)\notag\\
 &\phantom{q(t)=}-\frac{\varepsilon^3\, J_0^{3/2} }{2\, \sqrt{2}\, \omega_0^3}\,\cos (2\, \omega\,t)  +\frac{125\, \varepsilon^3\, J_0^{3/2} }{144 \,\sqrt{2}\, \omega_0^3}\,\cos (3 \omega\,t)-\frac{\sqrt{2}\, \varepsilon^3\,
   J_0^{3/2} }{3 \,\omega_0^3}\,\cos (4\, \omega\,t)+\frac{\varepsilon^3\, J_0^{3/2} }{16\, \sqrt{2}
\,   \omega_0^3}\,\cos (5 \,\omega\,t)\notag\\
&\phantom{q(t)=}+\frac{\varepsilon^3\, J_0^{3/2} }{18 \,\sqrt{2} \,\omega_0^3}\,\cos (6\, \omega\,t)-\frac{\varepsilon^3\, J_0^{3/2} }{48\, \sqrt{2}\, \omega_0^3}\,\cos(7\, \omega\,t)+\frac{\varepsilon^3\, J_0^{3/2} }{1296\, \sqrt{2}\, \omega_0^3}\,\cos (9 \,\omega\,t)\, ,\notag\\
&p(t)=\frac{\varepsilon\, J_0^{3/2} }{\sqrt{2}\, \omega_0}\,\sin (\omega\,t)-\frac{\varepsilon\, J_0^{3/2} }{3 \,\sqrt{2}
 \,  \omega_0}\,\sin (3\, \omega\,t)+\frac{5\,
   \varepsilon^2 \,J_0^2}{6\, \omega_0^2}-\frac{2\, \varepsilon^2\, J_0^2 }{3 \,\omega_0^2}\,\cos (\omega\,t)-\frac{2 \,\varepsilon^2\, J_0^2 }{3 \,\omega_0^2}\,\cos (2\, \omega\,t)\notag\\
   &\phantom{p(t)=}+\frac{2\,  \varepsilon^2\, J_0^2}{3 \,\omega_0^2}\, \cos (3\, \omega\,t)-\frac{\varepsilon^2\, J_0^2 }{6\, \omega_0^2}\,\cos (4\, \omega\,t)+\frac{31\, \varepsilon^3\, J_0^{5/2} }{12\, \sqrt{2}\, \omega_0^3}\,\sin (\omega\,t)-\frac{8\, \sqrt{2}\, \varepsilon^3\, J_0^{5/2} }{9\, \omega_0^3}\,\sin (2\,\omega\,t)\notag\\
  &\phantom{p(t)=}+\frac{97 \,\varepsilon^3\, J_0^{5/2} }{72\, \sqrt{2}\, \omega_0^3}\,\sin (3\, \omega\,t)-\frac{4 \,\sqrt{2}\, \varepsilon^3\,  J_0^{5/2} }{9 \,\omega_0^3}\,\sin (4 \,\omega\,t)+\frac{7\, \varepsilon^3\, J_0^{5/2} }{72\, \sqrt{2}\, \omega_0^3}\,\sin (5\, \omega\,t)\, ,\notag\\
&\omega=\omega_0-\frac{5\,\varepsilon^2\,J_0}{6\,\omega_0}\, .
\end{align}

\section{Proof of Lemma 1}
\label{appendixB}
Consider the Hamiltonian:
$$
\Hscr(q, J)=\omega_0\,J+\varepsilon\,h(q,J)\, ,
$$
where $h=\Oscr(J^{\frac{k}{2}})\,$ $k\in\mathbb{N}$, $k\geq 3\,$.  Introducing the translation $J=J_0+p\,$, $h=\Oscr((J_0+p)^{\frac{k}{2}})\,$, if $k$ is even we obtain a finite expression in the powers of $p$. If $k$ is odd, define $f(p)=\varepsilon\,(J_0+p)^{\frac{k}{2}}\,$ and suppose we want to compute the normalization steps by the Kolmogorov algorithm up to order $\varepsilon^n\,$ $n\geq 1\,$. The expansion in $p=0\,$ up to order $n$ yields
\begin{align*}
f(p)=\varepsilon\,(J_0+p)^{\frac{k}{2}}=\varepsilon\,J_0^{\frac{k}{2}}+\varepsilon\,\frac{k}{2}\,J_0^{\frac{k}{2}-1}p+\ldots+\frac{\varepsilon}{n!}\,\frac{k}{2}\left(\frac{k}{2}-1\right)\cdots\left(\frac{k}{2}-n+1\right)J_0^{\frac{k}{2}-n}p^{n}\, .
\end{align*} 
Moreover, observe that $\chi_{1}^{(1)} \sim \varepsilon\,J_{0}^{\frac{k}{2}}$ and $\chi_{2}^{(1)} \sim \varepsilon\, J_0^{\frac{k}{2}-1}\,$. Now, we identify those terms whose Lie derivatives could modify the generating functions along the normalization process. We have $L_{\varepsilon\chi_{1}^{(1)}}^{n-1}\left(\varepsilon \,J_0^{\frac{k}{2}-n}p^n\right)$. Thus the term of higher degree in $p$ in the expansion of $f(p)\,$, denoted by $\simeq f^{(n)}(p)\,$, influences $h_{n,1}^{(1)}$. In fact, since
\begin{align}
\label{grado.chi2}
L_{\varepsilon\chi_{1}^{(1)}}^{n-1}\left(\varepsilon \,J_0^{\frac{k}{2}-n}p^n\right)&=L_{\varepsilon\chi_{1}^{(1)}}^{n-2}\underbrace{\poisson{\varepsilon \,J_0^{\frac{k}{2}-n}p^n}{ \varepsilon\chi_1^{(1)}(q)}}_{\sim\,  \varepsilon^2 J_0^{k-n}\,p^{n-1}}=L_{\varepsilon\chi_{1}^{(1)}}^{n-3}\underbrace{\poisson{\poisson{\varepsilon \,J_0^{\frac{k}{2}-n}p^n}{ \varepsilon\chi_1^{(1)}(q)}}{ \varepsilon\chi_1^{(1)}(q)}}_{\sim\,  \varepsilon^3 J_0^{\frac{3}{2}\,k-n}\,p^{n-2}}\simeq\underbrace{\ldots\ldots}_{(n-3)\mathrm{times}}\\
&\simeq\, \varepsilon^n J_0^{\frac{k}{2}-n}\left(J_0^{\frac{k}{2}}\right)^{n-1} p = \varepsilon^n J_0^{\frac{(k-2)n}{2}}\, p\, \notag,
\end{align}
after $n$ step a contribution stemming from the term $f^{(n)}(p)$ will appear in the generating function $\chi_{2}^{(n)}\,$. On the contrary, the term $L_{\varepsilon\chi_{1}^{(1)}}^{n-1}\left(\varepsilon \,f^{(n+1)}(p)\right)$ contributes to $h_{n, 2}^{(1)}$ (which is quadratic in the action $p$), while the term $L_{\varepsilon\chi_{1}^{(1)}}^{n}\left(\varepsilon \,f^{(n+1)}(p)\right)$ contributes to $h_{n+1, 1}^{(1)}$, which is of order $\varepsilon^{n+1}$. From~\eqref{grado.chi2} it follows that $h_{n,1}^{(n)}=\Oscr\left(J_0^{(k-2)n/2}\right)\,$. Hence, the dependence of $\chi_2^{(n)}$ on the parameter $J_0\,$ is in the power $\chi_2^{(n)}\sim\, J_0^{\frac{(k-2)n}{2}}\,$. Also 
\begin{align*}
L_{\varepsilon\chi_{2}^{(1)}}^{n-1}\left(\varepsilon \,J_0^{\frac{k}{2}-1}p\right)&=L_{\varepsilon\chi_{2}^{(1)}}^{n-2}\underbrace{\poisson{\varepsilon \,J_0^{\frac{k}{2}-1}p}{ \varepsilon\chi_2^{(1)}(q, p)}}_{\sim\,  \varepsilon^2 J_0^{k-2}\,p}=\underbrace{\ldots\ldots}_{(n-2)\mathrm{times}}\simeq\, \varepsilon^n J_0^{\frac{k}{2}-1}\left(J_0^{\frac{k}{2}-1}\right)^{n-1} p = \varepsilon^n J_0^{\frac{(k-2)n}{2}}\, p\, ,
\end{align*}
thus, again, $h_{n,1}^{(n)}=\Oscr\left(J_0^{(k-2)n/2}\right)\,$. Finally
\begin{align*}
L_{\varepsilon\chi_{1}^{(1)}}^{n-1}\left(\varepsilon \,J_0^{\frac{k}{2}-n+1}p^{n-1}\right)&=L_{\varepsilon\chi_{1}^{(1)}}^{n-2}\underbrace{\poisson{\varepsilon \,J_0^{\frac{k}{2}-n+1}p^{n-1}}{ \varepsilon\chi_1^{(1)}(q)}}_{\sim\,  \varepsilon^2 J_0^{k-n+1}\,p^{n-2}}=
\underbrace{\ldots\ldots}_{(n-2)\mathrm{times}}\simeq\, \varepsilon^n J_0^{\frac{k}{2}-n+1}\left(J_0^{\frac{k}{2}}\right)^{n-1} p \\
&= \varepsilon^n J_0^{\frac{(k-2)n+2}{2}}\, p\, ,
\end{align*}
implying  $\chi_1^{(n)}\sim\, J_0^{\frac{(k-2)n+2}{2}}\,$.
This concludes the proof of the lemma. \qed

\end{document}